\documentclass[final, authoryear, 12pt]{elsarticle}

\usepackage{amssymb}

\usepackage{threeparttablex}
\usepackage{longtable}
\usepackage{geometry}
\usepackage{tabularray}
\usepackage{array}
\usepackage{hyperref}
\usepackage{makecell}
\usepackage{booktabs}
\usepackage[nolist]{acronym} 
\usepackage{xcolor}
\usepackage[absolute,overlay]{textpos}
\usepackage{boxedminipage}

\usepackage[export]{adjustbox}

\journal{Sustainable Cities And Society}

\begin{document}

\begin{frontmatter}



\begin{textblock*}{\textwidth}(3cm, 1cm)
\begin{center}
\begin{footnotesize}
\begin{boxedminipage}{1\textwidth}
This is the Accepted Manuscript version of an article published by Elsevier in the journal \emph{Sustainable Cities and Society} in 2025, which is available at:\\ \url{https://doi.org/10.1016/j.scs.2025.106441}\\ Cite as:
Cai, C., Li, B., Zhang, Q., Wang, X., Biljecki, F., Herthogs, P. (2025). Bi-directional mapping of morphology metrics and 3D city blocks for enhanced characterisation and generation of urban form.  \textit{Sustainable Cities and Society}, 106441.
\end{boxedminipage}
\end{footnotesize}
\end{center}
\end{textblock*}

\vspace*{2cm}

\title{Bi-directional Mapping of Morphology Metrics and 3D City Blocks for Enhanced Characterisation and Generation of Urban Form} 

\author[sec]{Chenyi Cai \corref{cor1}}
\cortext[cor1]{Corresponding author: Chenyi Cai. Email: \texttt{chenyi.cai@sec.ethz.ch}}

\affiliation[sec]{organization={Singapore-ETH Centre, Future Cities Lab Global Programme},
addressline={CREATE campus, 1 Create Way, \#06-01 CREATE Tower}, 
            postcode={138602}, 
            country={Singapore}}
            
\author[seu]{Biao Li}
\affiliation[seu]{organization={School of Architecture, Southeast University},
            addressline={2 Sipailou}, 
            city={Nanjing},
            postcode={210096}, 
            country={China}}
            
\author[ethz]{Qiyan Zhang}
\affiliation[ethz]{organization={Department of Architecture, Swiss Federal Institute of Technology Zurich},
            addressline={Stefano-Franscini-Platz 1}, 
            city={Zürich},
            postcode={8093}, 
            country={Switzerland}}
            
\author[seu]{Xiao Wang}
            
\author[nus,dre]{Filip Biljecki}
\affiliation[nus]{organization={Department of Architecture, National University of Singapore},
            addressline={4 Architecture Drive}, 
            city={Singapore},
            postcode={117566}, 
            country={Singapore}}
\affiliation[dre]{organization={Department of Real Estate, National University of Singapore},
            addressline={15 Kent Ridge Drive}, 
            city={Singapore},
            postcode={119245}, 
            country={Singapore}}
            
\author[sec]{Pieter Herthogs}

\begin{abstract}
Urban morphology examines city spatial configurations and plays a pivotal role in urban design and sustainability.
Morphology metrics are essential for performance-driven computational urban design (CUD), which integrates the automatic generation of urban form and the evaluation and optimisation of urban performance.
Although form generation and performance evaluation both rely on morphology metrics (e.g., floor area ratio), they are rarely unified into one workflow.
Typically, form generation methods follow one-directional metric-to-form logic, whereas performance evaluation methods adopt the inverse form-to-metric logic.
As a result, morphology metrics are often used in isolation within each method, limiting their applicability across both processes.
To address this gap, approaches that can support bi-directional workflows, namely, simultaneous form-to-metric and metric--to--form, have the potential to combine and exchange results from both sides.
The methodology introduced in this paper, which we refer to as bi-directional mapping, enables the formulation of sets of morphology metrics derived from form and then enable metric-to-form translation.
We present approaches to formulate metric sets composed of indicators related to urban form and performance to characterise complex urban form and support performance evaluation.
The metric sets can be derived from different cities, with 3D urban models of New York City as a demonstration in this study.
Artificial neural networks are used to cluster 3D models and encode morphology metrics, enabling the generation of diverse urban models through case retrieval.
Additionally, the effectiveness of the metrics in representing 3D city blocks is evaluated through comparative analysis.
Our methodology identified metric sets that can comprehensively characterise 3D city blocks and enable effective retrieval for generating similar urban models. This improves performance-driven CUD towards sustainable urban design and planning.

\end{abstract}








\end{frontmatter}

\begin{acronym}
 \acro{CUD}{Computional Urban Design}
 \acro{SOM}{Self-organising map}
 \acro{UMI}{Urban morphology indicator}
 \acro{OSM}{Open Street Map}
 \acro{PCC}{Pearson correlation coefficient}
 \acro{BMU}{Best matching unit}
\end{acronym}


\section{Introduction}
\label{sec:intro}
Urban form, or the spatial structure of cities, is typically examined as the physical spatial configuration within urban spaces \citep{lynch1958theoryofurbanform, kropfAspectsUrbanForm2009, chiaradia2019urbanmophoandform}. 
Urban morphology, the study of urban form, focuses on how the physical form and structure of cities are shaped and organised. It examines the patterns and layouts of buildings, open spaces, streets, and other elements that make up the built environment \citep{moudon1997urbanmorphologyasemerging, scheer2016epistemologyofUM, kropf2018handbook, ratti2003buildingformenvironmentalperformance}. 
Urban morphology studies support the evaluation of urban performance, such as microclimate, heat and energy, mobility resilience, and the entire socio-economic and technological fabric of urban systems \citep{zhou2022understandingUMeffectonLST, mashhoodi2024urbanformandLST, rode2014citiesandenergy, lee2016impactofindividualtraits, bramley2009urbanformandsustainability}. 

Morphology metrics, quantifiable representations of characteristics of urban form, are fundamental to understanding the urban morphology of the built environment and enabling urban form generation and performance optimisation in computational urban design (CUD) \citep{dibble2019originofspacesmorphometrics, zhang2023spatialmeasures}. 
Morphology metrics that are widely used in urban form studies, such as floor area ratio, sky view factor, and others, have been used to assess solar and energy efficiency \citep{chatzipoulka2018skyviewfactorforsolaravailability}. 
Researchers use different sets of morphology metrics to investigate the relationship between urban form and sustainability outcomes, such as urban ventilation, air pollution, urban heat island (UHI), transportation mode and more \citep{galster2001wrestlingsprawl, li2022exploringurbanventilation, liu2017patternsofairpollution, yin2018urbanformandUHI, rybarczyk2014urbanformbicyclemode, zhang2023spatialmeasures}. 
Consequently, morphology metrics are essential in automatic generation of urban form and evaluation and optimisation of urban performance, which are the three main components of performance-driven computational urban design (CUD) \citep{shi2017reviewsimulationbasedurbanformgeneration,koenig2020integratinganalysisanddesign,zhang2024reviewurbanformgenerationandoptimization}.

\begin{figure}[h]
\includegraphics[width=.95\textwidth]{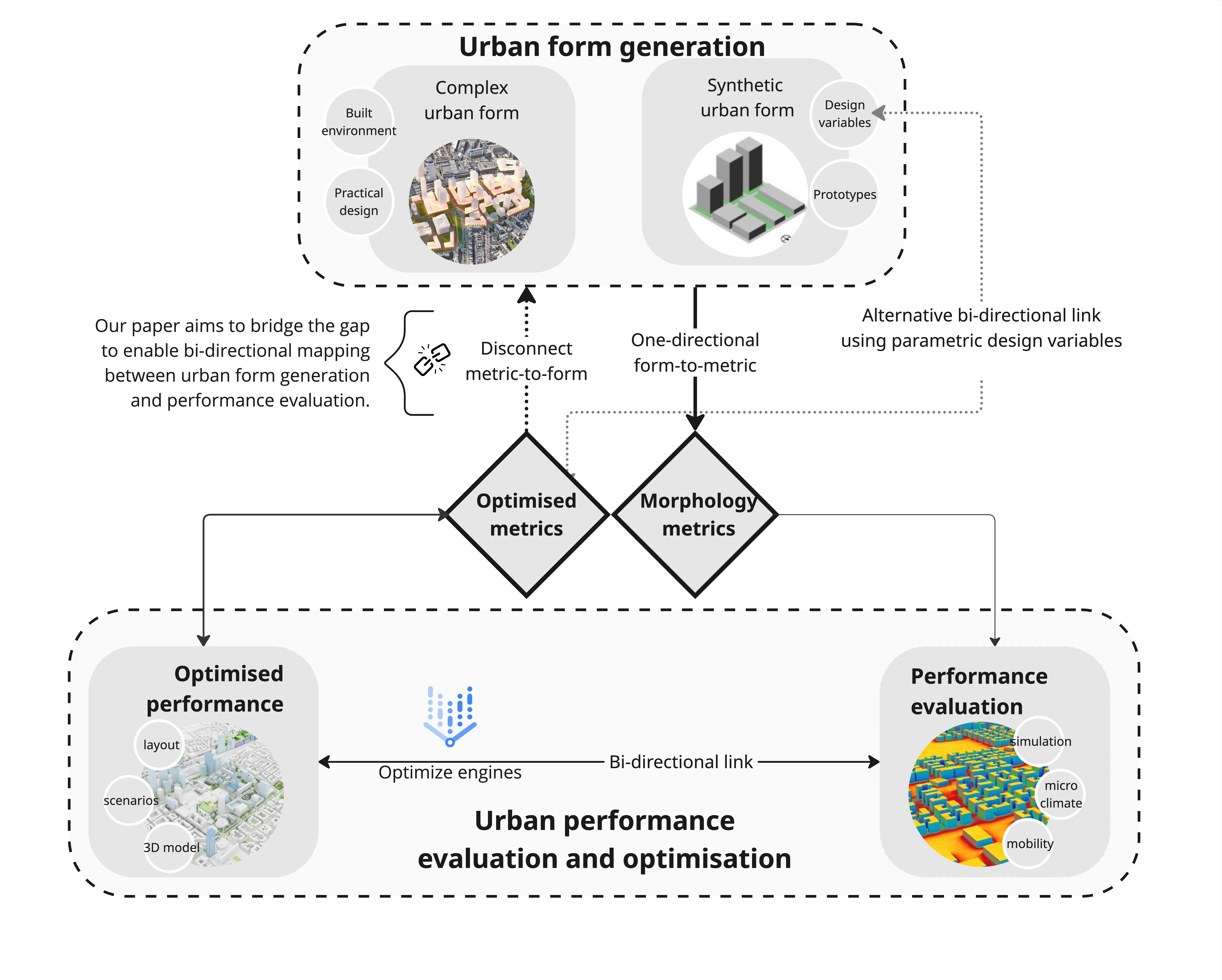}
\centering
\caption{The general workflow of performance-driven computational urban design includes urban form generation, performance evaluation and optimisation. It shows the location of the disconnection between performance evaluation and urban form generation.}
\label{fig:problemstatement}
\end{figure}

Although form generation and performance evaluation both rely on morphology metrics, they are rarely unified into one workflow. Typically, in performance-driven CUD workflow, form generation methods follow one-directional metric-to-form logic, whereas performance evaluation methods adopt the inverse form-to-metric logic. As a result, morphology metrics are often used in isolation within each method, limiting their applicability across both processes. 
A key challenge is translating the morphology metrics into diverse urban form, especially in the performance evaluation process. 
Consequently, the absence of the metric-to-form workflow hinders translating optimised morphology metrics derived from the performance evaluation solution space to the generation of improved complex urban form (see Figure \ref{fig:problemstatement} the disconnect metric-to-form direction).
As shown in Figure \ref{fig:problemstatement}, ideally the performance-driven CUD process should operate as multiple iterations of exchanging results between metric and form. 
One iteration begins with the initial urban form, followed by derive morphology metrics from the form, then followed performance evaluation based on its morphology metrics.
Then optimised morphology metrics are identified through optimisation engines to enhance performance. 
These optimised metrics are subsequently used to generate improved urban forms, enabling iterative refinement and continuous integration between performance evaluation, optimisation, and form generation.
However, when optimised morphology metrics cannot be translated to generate corresponding 3D urban form models --- primarily the absence of the metric-to-form workflow --- it becomes challenging to produce urban form models based on these optimised metrics.

Currently, two main approaches generally attempted to address the gap mentioned above.
The first approach involves using parametric design variables as alternative representations of urban form, which facilitates the generation of urban form (see Figure \ref{fig:problemstatement}).
Several works manipulate the urban form using design variables from simplified geometries or predefined prototypes to generate synthetic urban forms \citep{panao2008optimizationurbanbuilding, vermeulen2013urbanformoptimization,kampf2010optimisationofbuildingform,gagne2012generativefacadedesign, yi2015agentbasedgeometryoptimization, zhang2019impacttopologysolarpotential}. 
The corresponding morphology metrics are subsequently calculated from these generated forms and used for performance evaluation. Through iterative adjustments of the design variables and repeated execution of this process, optimised urban form is identified \citep{panao2008optimizationurbanbuilding, vermeulen2013urbanformoptimization,gagne2012generativefacadedesign, yi2015agentbasedgeometryoptimization, choi2021designofurbantissueevolutionary}.
However, while this method produces an optimised urban form driven by performance, the generated urban form is limited to simplified prototypes, which is insufficient to capture the complexity of real-world urban environments. 
The second approach addresses the complexity of urban form by incorporating existing cases from the built environment or urban design proposals. 
In this method, morphology metrics related to urban performance are calculated from existing models, allowing the performance evaluation of the urban form studied\citep{yu2015applicationofmulti-objectiveGA,kampf2009optimisationofurbanenergy,rode2014citiesandenergy}. 
However, while this approach can deliver optimised metrics that can enhance urban performance, it does not facilitate the generation of optimised urban models based on these metrics. As a result, it limits the automation of the whole process.

The two primary approaches still present a trade-off: they can achieve either an automated optimisation process or urban form complexity, but not both simultaneously. 
They have provided workarounds to address the gap; however, they do not fully resolve it.
The work presented in this paper attempts to address this gap. 
The gap arises from the fact that, in performance-driven CUD, while morphology metrics are used for performance evaluation and optimisation, the optimised metrics cannot be effectively applied to generate corresponding improved urban form models. 
We argue that the gap fundamentally stems from the absence of metric-to-form approaches to translate morphology metrics into diverse urban form.
Therefore, it is crucial to develop systematic methods which has the capacity to support bi-directional, metric-to-form and form-to-metric logics to combine form generation and performance evaluation methods and exchange results between them.
While quantifying 3D urban form is more straightforward when morphology metrics are identified, by further establishing a robust metric-to-form connection, it enables a true bi-directional mapping -- both form-to-metric and metric-to-form processes.
This bi-directional mapping can ultimately bridge the gap between performance optimisation and urban form generation, advancing computational urban design.

To establish the bi-directional mapping methodology between morphology metrics and urban form, two aspects are essential:1) formulating metric-to-form where morphology metrics can both characterise complex urban form and support performance evaluation, and 2) developing a systematic methodology to enable metric-to-form workflow.
Approaches are needed to both formulate these metrics and evaluate their effectiveness in representing urban morphology, particularly for 3D urban models. 
More importantly, the approaches must enable form generation based on input morphology metrics to align with the desired urban form types. They should also address the complexity of urban forms by generating 3D models that are consistent with the intended types and exhibit sufficient richness and diversity.

A set of morphology metrics, which is a combination of multiple urban morphology indicators (UMI), should be a meaningful measurement of urban form, depicting shape complexity, measuring relative richness and diversity, and quantifying aggregation and contagion \citep{dibble2019originofspacesmorphometrics,O1988indicesoflandscapepattern}. 
The complexity of the characteristics of the urban form cannot be encapsulated in a single morphology indicator nor in the sheer number and diversity of indicators. 
There are UMIs that describe various urban units, including streets, plots, buildings, and open spaces \citep{elzeni2022classificationofUM}. They vary in scales including landscape, surface, urban canopy, and buildings \citep{liMultidisciplinaryParametersCharacterizing2024, biljeckiGlobalBuildingMorphology2022}.
Therefore, further investigations are needed to formulate effective morphology metric sets for complex urban form characterisation.

A city block is an important unit for understanding and interpreting urban design. 
City blocks are the spaces for buildings within the street pattern of a city and form the basic unit of a city's urban fabric. 
The characterisation of 2D urban morphology has been extensively investigated \citep{labetski20233dbuildingmetrics, biljeckiGlobalBuildingMorphology2022} and applied in topics such as changes in urban pattern \citep{herold2002structureandchangesoflanduse} and regionalisation \citep{yang2022urbanmorphologicalregionalization}, extraction of morphological features \citep{cai2021urbanmorphologyfeatureextraction} and urban form-energy relations \citep{mashhoodi2024urbanformandLST}.
However, the morphology metrics evaluation methods in representing such block-scale building combinations, particularly 3D models, have not been fully discovered. 
The urban morphology on the block scale contains a group of buildings rather than a single architectural element \citep{berghauserpontSpacematrixSpaceDensity2021a} and the characteristics of the block environment have significant effects on energy efficiency \citep{wan2024exploringblockenvironmentalcharacteristics}. 
Beyond 2D information, the vertical dimensions playes a crucial role in comprehensive urban morphology representation \citep{yang2022urbanmorphologicalregionalization}.
Therefore, we need approaches to evaluate the effectiveness of morphology metrics in representing the morphology of 3D block-scale buildings. These approaches need to be flexible and adaptable when applied to urban blocks of different regions.

Clustering --- the unsupervised machine learning technique --- is a data-driven approach that has been widely applied to analyse spatial data and urban form, improving our understanding of city structures.
By classifying patterns in urban form, clustering methods can help identify different built environments, support urban planning, and inform sustainable design strategies \citep{li2024characterizing}.
Researchers have explored various clustering techniques for urban form analysis, including block-scale morphology and 3D building structures \citep{qu2023blocks,cai2022dataclusteringinUCM,li2023identifyingtypologies,zhao2023influenceofUMonfacadesolarpotential}.
Although most urban form clustering studies are descriptive and analytical, a few research have applied clustering approaches to differentiate 3D morphology and enable case retrieval based on morphological similarity, but limited in individual buildings \citep{labetski20233dbuildingmetrics}. 
Clustering is effective in classifying urban form and has the potential to retrieve and generate diverse block-scale 3D models based on abstract morphology metrics.

We propose systematic methodology that addresses the challenge of translating morphology metrics into diverse urban models, through formulating effective morphology metrics and establishing bi-directional mapping between the metrics and form.
Our systematic workflow can derive city block morphology metrics from 3D models to simultaneously improve the characterisation, generation, and performance evaluation for urban form. 
Meanwhile, our approaches can evaluate the effectiveness of morphology metrics in representing form characteristics through comparisons. 
We demonstrate it with 3D urban data from New York City, including 14\,248 blocks. The self-organizing map (SOM) is used to cluster and encode the block-scale 3D models. 
Allowing the flexible retrieval of diverse 3D models according to the input morphology metrics of the target urban form, enhance the bi-directional mapping morphology metrics and 3D urban form models.
In this study, we identified effective morphology metrics to characterise 3D city blocks in our data set by comparing multiple sets. 
As a result, our identified morphology metrics not only effectively represent the complexity of block-scale 3D urban models, but also support the generation of diverse 3D models that align with intended urban form types. 
Additionally, the methodology is generalisable, allowing the adaptations and applications across varied contexts.
The approaches can further enable the generation of improved urban form models through optimised morphology metrics linked to improved performance.

The remainder of the paper is structured as follows. Section \ref{sec:bg} provides an overview of the literature on existing urban morphology metrics, their applications in the generation and evaluation of urban form, and theinherent challenges. Section \ref{sec:method} presents our methodology for the construction of 3D block-scale urban morphology indicators and clustering based on metric sets. Section \ref{sec:result} demonstrates how our methodology can be applied for retrieving 3D urban models based on morphology metrics derived from provided urban form, as well as evaluating the effectiveness of the metric sets. Finally, in sections \ref{sec: discuss} and \ref{sec:conclusion}, we summarise key contributions, acknowledge remaining limitations, and outline directions for future work.

\section{Background and related work}
\label{sec:bg}

\subsection{Performance-driven urban form generation}
Urban form, or the spatial structure of cities, is typically examined as the physical configuration of various elements within urban spaces \citep{batty1976urbanModeling, kropf2018handbook,lynch1958theoryofurbanform}. 
Urban form demonstrates significant correlations with urban performance, for instance, it impacts micro-climates, thermal and energy dynamics, mobility resilience, and the broader socio-economic-technological structure of urban systems \citep{anderson1996urbanformenergyandenviron,chiaradia2019urbanmophoandform,zhou2022understandingUMeffectonLST,lee2016impactofindividualtraits, bramley2009urbanformandsustainability}. 
Urban form generation, one of the main topics in computational urban design, provides systematic approaches to produce design solutions and accelerate urban design decisions and processes by using computational tools \citep{batty1991generatingurbanforms, jiang2024generativeurbandesignreview, wilson2019generatemasterplans}. The urban form generation process involves modelling methods of representing and quantifying urban forms to be machine-readable and computable, and optimization methods to evolve the urban forms to meet pre-defined goals \citep{miao2018computationalurbandesignprototyping, jiang2024generativeurbandesignreview}. 

Quantified measurements of urban forms, which are associated with urban form characteristics, are essential for performance-driven urban form generation, especially for urban environmental performance evaluation and optimisation. 
One of the advantages of performance-driven CUD workflow is to enable the automatic generation of sustainable urban spatial configurations by optimising design solutions based on environmental metrics \citep{koenig2020integratinganalysisanddesign, canuto2024predictive-performativedesign}. Researchers have developed various computational and parametric computer-aided design tools for urban form generation, performance evaluation, and performance-driven optimisation \citep{shi2017reviewsimulationbasedurbanformgeneration, zhang2024reviewurbanformgenerationandoptimization}. 
During the process, researchers use a variety of urban morphology metrics to measure the characteristics of urban form in different aspects and contexts \citep{zhang2023spatialmeasures}.
The relationships between urban form and environmental performance have been investigated from various perspectives. Morphological metrics have proven effective in analyzing sustainable outcomes.
A broad spectrum of research has leveraged urban morphology metrics, such as site coverage \citep{li2024high-densityurbanblockformgeneration}, sky view factor \citep{ratti2003buildingformenvironmentalperformance, li2022exploringurbanventilation}, built volume \citep{rode2014citiesandenergy}, building height \citep{yi2015agentbasedgeometryoptimization,li2024high-densityurbanblockformgeneration}, height-to-width ratio \citep{elzeni2022classificationofUM}, and others to analyze, predict, and optimize environmental performance metrics such as walkability \citep{rakha2012urbanmodelingwalkability}, thermal comfort \citep{yu2015applicationofmulti-objectiveGA}, solar efficiency \citep{rode2014citiesandenergy,li2024high-densityurbanblockformgeneration}, urban ventilation \citep{li2022exploringurbanventilation} and so on.
Urban morphology metrics are central to performance-driven urban design as they serve as a crucial link between urban forms and performance evaluation and optimisation. 

In performance-driven computational urban design, the connections between urban form to morphology metrics are mostly form-to-metric one-way links, hindering the fully automated process of performance-driven urban form generation, as illustrated in Figure \ref{fig:problemstatement}. 
The gap is typically addressed through two main approaches. 

Firstly, researchers incorporate parametric approaches for manipulating urban forms. Design variables are identified according to predefined prototypes which are simplified geometries to generate synthetic urban models. 
An example of replicating a specific urban form is to simulate morphology, and building structure form using shape grammar, translating the properties and their interrelationships to geometric rules \citep{wang2020generativeurbandesignshapegrammar,mandic2015shapegrammarapplicationinUD}. 
\cite{li2024high-densityurbanblockformgeneration} generated simplified block-scale urban forms based on grids and linked solar performance with urban form metrics such as floor area ratio (FAR) and site coverage (SC). It enables the integration of form generation, solar simulation, and optimisation for multi-objective solar performance. 
\cite{shen2024environmentalperformancedrivenurbanmodularhousing} utilised cellular automata for synthesising simplified layouts that compiled 3D grids, enabling a form generation process that can be integrated with daylight performance simulation and generic optimisation to produce housing layouts.
\cite{zhang2019impacttopologysolarpotential} focused on generating blocks based on six typology categories (e.g. blocks with towers, courtyards), and then morphological metrics (e.g. Open Space Ratio (OSR), Roof-to-envelope area ratio) derived from these synthesized blocks are used to identify the impacts of design factors on solar potential and energy performance \citep{zhang2019impacttopologysolarpotential}.
The process incorporates optimisation methods whereby iterative adjustments of design variables and repeated execution yield a solution space that aligns with predefined performance objectives, thereby contributing to the generation of optimised urban forms \citep{koenig2020integratinganalysisanddesign, stouffs2015generativeandevolutionary}.
However, while this method produces performance-driven optimised urban models, these models are limited to simplified prototypes, insufficient to capture the complexity of real-world urban environments. 

Secondly, the other approach addresses the urban form complexity by utilising cases from the built environment, where the urban form is more realistic and complex with diverse variants. 
\cite{zhou2022understandingUMeffectonLST} extracted morphological metrics such as sky view factor (SVF) and building surface fraction (BSF), from the built environment. Research shows that morphological metrics significantly affect land surface temperature (LST) at the local scale. 
\cite{li2022exploringurbanventilation} assessed the outdoor ventilation conditions that can be tightly correlated with the sky view factor (SVF). 
Indicators like floor area ratio (FAR), building density (BD), and street network density provide insights into land use efficiency and urban sprawl \citep{tsai2005quantifyingurbanform}. 
Through an analysis of the built environment in Nanjing, \cite{li2025multi-scale} examined eight morphological indicators,  such as floor area ratio (FAR), block surface ratio (BSR), alongside year-round solar radiation simulations. Their findings revealed strong correlations between morphological indicators and solar performance metrics.
\citep{zhou2024multi-oboptimisation} employed multi-objective optimisation approaches based on morphology indicators derived from the built environment, generating optimised performance outcomes (e.g., surface temperature, humidity). 
However, their work was limited to descriptive recommendations for morphological improvements, rather than producing optimised urban form.
Another similar study is from \cite{yu2015applicationofmulti-objectiveGA}, where building layouts were used in a multi-objective optimisation model for green building design. While the study identified optimal energy consumption levels and improved indoor thermal comfort, alongside key building design variables (e.g., floor area, window–wall ratio), it did not extend to generating optimised building layouts.

Similar challenges have been encountered in other studies attempting to link morphological parameters to the studied 3D urban forms, despite the availability of various optimisation methods (e.g. \cite{rode2014citiesandenergy, kampf2009optimisationofurbanenergy, zhang2024reviewurbanformgenerationandoptimization}). 
Consequently, investigations using complex real-world urban form can yield insights into optimised morphological parameters with enhanced performance outcomes, but lack providing improved urban forms that can directly inform design decisions.
Therefore, to support the automated performance-driven urban form generation, it is essential to formulate morphology metrics that can both effectively characterise urban forms and enable the metrics-to-form generations.

\subsection{Urban morphology metrics}
Multiple representations, quantifications and parameters are utilised towards computational design for modelling and simulating \citep{batty1976urbanModeling,jiang2024generativeurbandesignreview}. As the main components of the 3D urban morphology, buildings are arranged in various configurations throughout cities, creating diverse morphologies \citep{liu2014impactsof3DUM,liu2020characterizing3dresidential}. 
Each building possesses distinct physical attributes, such as height, area, volume, and outline. These unique characteristics collectively influence the overall 3D urban morphology of a region. 
Hence many efforts have been made to develop morphology metrics for characterizing, analyzing and understanding spatiotemporal patterns of the built environments. 

Multi-disciplinary investigations for quantitatively analysing and representing urban form characteristics vary in methods and scales \citep{liMultidisciplinaryParametersCharacterizing2024, zhang2023spatialmeasures}. 
Urban form can be quantified and represented by urban morphology metrics using measurements from various aspects -- distribution of buildings, streets, and open spaces. 
With the explosive growth of information technology and available multi-source data, researchers from multiple disciplines have developed various approaches for quantifying morphology, including shape grammar approaches, spatial analytical approaches, ML-based approaches and urban morphological metrics \citep{kropf2018handbook, biljeckiGlobalBuildingMorphology2022,liMultidisciplinaryParametersCharacterizing2024}. 
Researchers have investigated comprehensive lists of UMIs with multi-scale measures through systematic literature reviews \citep{biljeckiGlobalBuildingMorphology2022, liMultidisciplinaryParametersCharacterizing2024}.
Morphology metrics can represent the complexity, relative richness and diversity of urban forms. 
By employing a variety of spatial metrics, researchers can assess the compactness, connectivity, and complexity of urban environments, which in turn inform planning and policy decisions \citep{lowry2014comparingspatialmeatures}. 

Researchers have explored various approaches to linking morphology metrics with urban form. 
A range of morphology metric sets has been developed for urban form quantification, classification, clustering, and prototyping \citep{schirmer2016multiscale}, primarily following a form-to-metric method.
Traditional urban morphology metrics were widely used to quantify two-dimensional spatiotemporal patterns. 
\cite{berghauserpontSpacematrixSpaceDensity2021a} developed Spacemate with a focus on various types of density on the urban block such as compactness (FAR), the coverage (BD), spaciousness: open space ratio (OSR), network density and the average number of storeys \citep{berghauserpontSpacemateDensityTypomorphology2005a}. 
Spacemate is widely recognised in academic and research fields, particularly in urban form classification, density analysis, and typology comparisons. Yet it is mainly used to classify and understand urban form density, and its computational effectiveness has not been evaluated.
Vertical information in urban morphology has been underscored to advance three-dimensional spatial pattern analysis based on new data sources and techniques \citep{liu2020characterizing3dresidential}. 
For instance, OpenStreetMap (OSM) is an open digital mapping database created through crowdsourced and volunteer-contributed geographic information, including buildings and infrastructure worldwide~\citep{Herfort2023,2023_bae_osm_qa}. 
Researchers have advanced numerous 3D indicator systems based on established theories and practices \citep{liMultidisciplinaryParametersCharacterizing2024,labetski20233dbuildingmetrics}. 
Volumetric approaches have been used as a means of better capturing the 3D morphological characteristics of cities \citep{bruyns2021urbanvolumetrics}. 
Morphology metrics used are also defined and delineated in city planning regulations, and it has been demonstrated that one can generate 3D urban form from regulations \citep{chadzynski2022semantic3dcityagents, grisiute2023semanticspatialpolicy}.
Other scholars have developed 3D spatial metrics based on gradient surface models, focusing on metric development on buildings to quantify urban morphology \citep{kedron20193DSpatialMetricsForObjects}. 
Indicator sets are valuable for better understanding urban forms and for spatial analysis. However, the indicator sets remain empirical in their ability to describe complex urban forms, and it has yet to be validated whether they are sufficient for computers to effectively represent urban form characteristics.

Urban form clustering leverages the increasing availability of extensive spatial data to effectively define urban form characteristics and uncover knowledge for a better understanding of urban morphology \citep{li2024characterizing}.
While researchers have explored various clustering methods for urban form analysis, most studies remain descriptive or analytical, as the metric-to-form mapping --- the effective retrieval of diverse urban forms from abstract morphology metrics --- remains a significant challenge.
In 3D block form studies, \cite{qu2023blocks} proposed a 35D form index system for block morphology and applied t-SNE clustering to categorise urban blocks, subsequently assigning clustering results across the city. Their research concluded with a comparative analysis of block characteristics with urbanisation, socioeconomic development, and sociocultural factors. 
Similarly, \cite{li2024characterizing} applied clustering methods to 3D building data, identifying eight urban built form typologies with the aim of informing urban governance and planning practices.
Beyond analytical clustering studies, some research has attempted to retrieve urban forms based on similarity analysis. 
For example, \cite{xuRenewalDesignofRoma2019} used shape indicators to identify block footprint characteristics, developing a case-based method for an urban renewal project in Rome. However, their approach only addressed 2D block shapes and did not account for the 3D building forms within the blocks.
A very relevant study is that of \cite{labetski20233dbuildingmetrics}, which examined 3D building metrics to represent the distinct characteristics of complex 3D building shapes. One of their case studies involved hierarchical clustering to verify the effectiveness of 3D indices in differentiating building forms, assessing both the uniformity and distinctiveness of the clustered results. However, their study focused on single-building level characterisation rather than urban block-level morphology.

Based on the existing literature, we first observe a disconnection in the fully automated process of performance-driven urban form generation and optimization, caused by the lack of bi-directional mapping between morphology metrics and urban form, particularly in the metric-to-form direction.
There are approaches to deriving metrics from models for evaluating urban performance using these metrics.
However, there is a lack of approaches that couple metrics and form, especially generating urban form that go beyond analytical studies (e.g., simplified prototyping). 
The disconnection in generating improved 3D urban models based on optimized morphology metrics with optimized performance is the key issue.
Secondly, the complexity of urban form characteristics cannot be adequately represented by a single indicator or soley by quantity and variety of indicators. 
Consequently, identifying the urban morphology metrics that can capture aspects such as shape complexity, relative richness and diversity, for 3D urban models is important.
Thirdly, clustering-based approaches have the potential to uncover urban form complexity; however, the simultaneous evaluation of morphology metrics and the retrieval of block-scale 3D models have not been fully explored.
The following sections propose several contributions to achieve the goal of bridging morphology metrics and complex urban forms bi-directionally to enhance performance-driven CUD. 

\section{Methodology}
\label{sec:method}
\begin{figure}[h]
\includegraphics[width=\textwidth]{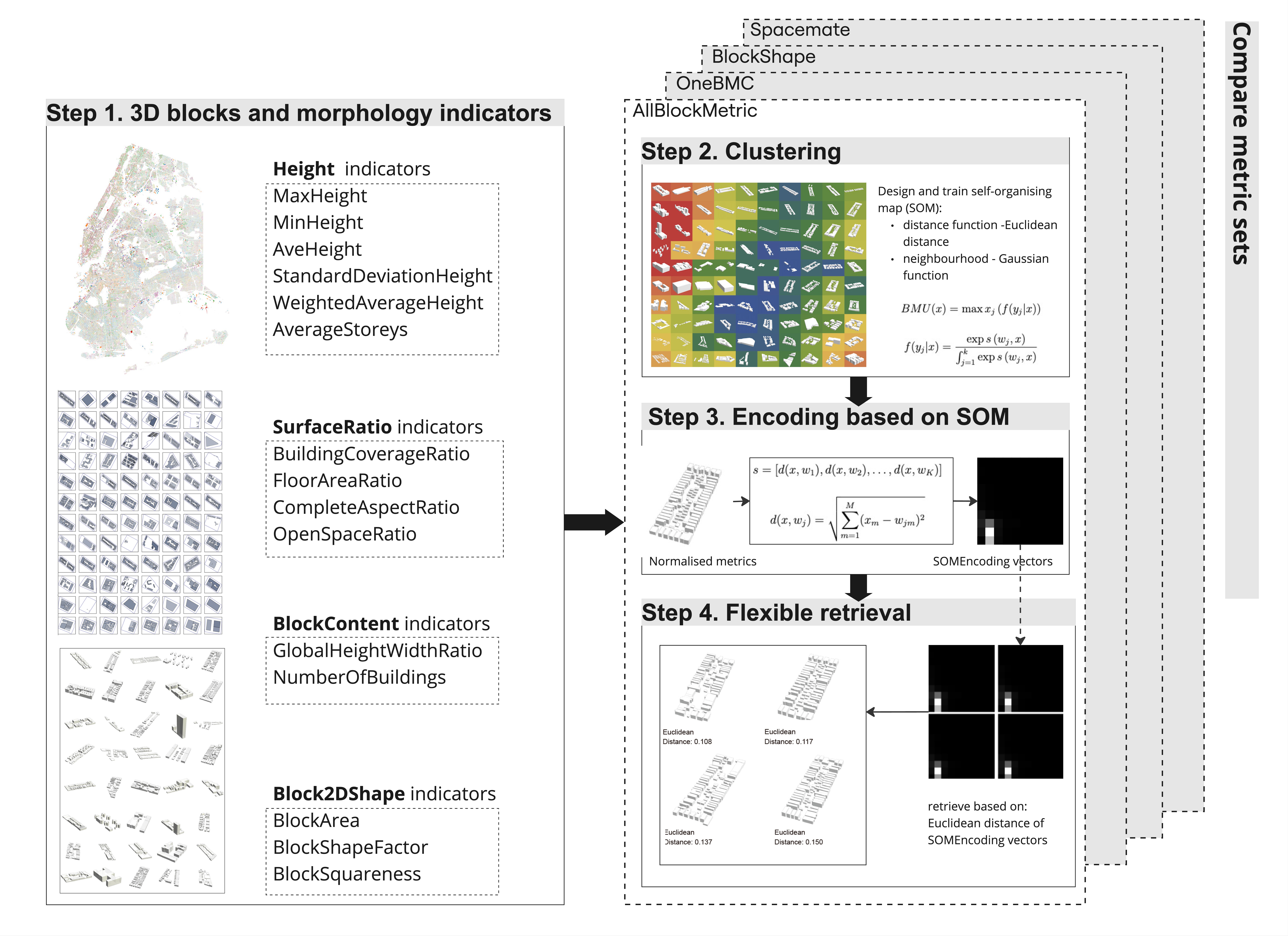}
\centering
\caption{Our proposed systematic framework include four main steps. First, the morphology metrics of 3D blocks incorporate four characteristics types with 15 indicators. Second, Self-organising maps are trained to cluster city blocks. Third, based on the trained SOM, a context-based encoding method is introduced to represent urban form, generating SOMEncoding vectors. Fourth, 3D models similar to the target are retrieved using SOMEncoding vectors. In this paper, four metric sets are designed to compare their effectiveness in characterising 3D city blocks.}
\label{fig:workflow}
\end{figure}

This section introduces our methodology for bridging morphology metrics and 3D urban form bi-directionally, enabling urban form generation based on morphology metrics derived from targeted 3D block-scale models. It is a systematic pipeline that includes data collection and pre-processing, morphology metrics formulation, clustering, encoding and case retrieval techniques. 
We summarise our methodology in Figure \ref{fig:workflow}.

We set out the implementations based on the following scopes and aims: 
\begin{itemize}
    \item In preference of open-source geoinformation, 3D urban form data were collected from readily accessible platforms. We used data from OpenStreetMap, where we sliced the blocks based on the provided road network categories. In our work, the effectiveness of morphology metrics is evaluated through block-scale urban form characteristics from New York City, as it is our chosen study area. 
    \item To develop effective morphology metrics, the selected UMIs are unambiguous, can be universally present in all cities and have geometric significance in representing block-scale 3D urban form characteristics of shape complexity. The morphology metrics we selected are primarily drawn from existing literature, which are related to 3D building morphology in blocks, so they are not exhaustive. However, they cover the four main urban form characteristic types: height, surface, content, and shape. 
    \item The effectiveness of morphology metrics is assessed through comparative evaluation. In the experiments comparing different sets of morphology metrics, a guiding principle is that the morphology metric set can represent the constituent elements of urban form such that their homologous characteristics can be rigorously defined and measured \citep{dibble2019originofspacesmorphometrics}. In our study, city blocks retrieved based on similarity in morphology metrics are expected to display analogous morphological characteristics while maintaining diverse form variations.
    \item The approaches for formulating morphology metrics and establishing the bi-directional mapping between these metrics and 3D urban form are designed to be generalisable to other cities with varying urban structures. We provide a flexible and context-based methodology that can be adaptively applied to other datasets which can be plot-wise or from other regions. Our study does not attempt to create a universal set of morphology metrics for all regions. When our proposed approaches are applied to other urban form datasets, additional UMIs can be incorporated and further examined according to specific contexts.
\end{itemize}

\subsection{Data collection and pre-processing}
\begin{figure}[h]
\includegraphics[width=\textwidth]{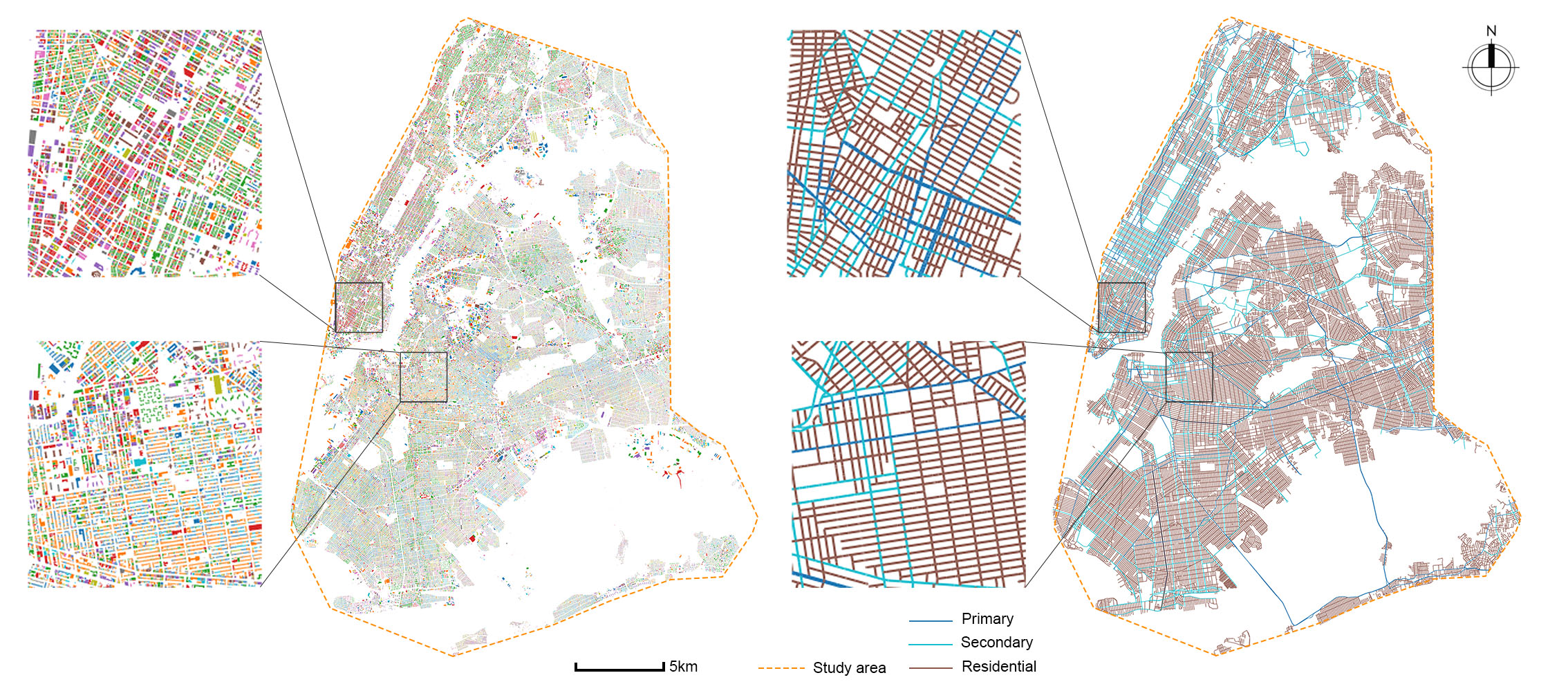}
\centering
\caption{Our studied area and collected data include building footprints (left) and road networks (right). (c) OpenStreetMap contributors.}
\label{fig:datacollection}
\end{figure}

An area of New York City was downloaded from OpenStreetMap, including building footprints and road networks(Figure \ref{fig:datacollection}). In the figure, buildings are coloured based on their height information. Roads are classified as Primary, Secondary, and Residential. Residential roads primarily define block-scale units; therefore, the urban blocks used for further studies were delineated based on the boundaries set by these three road types.

\begin{figure}[h!]
\includegraphics[width=\textwidth]{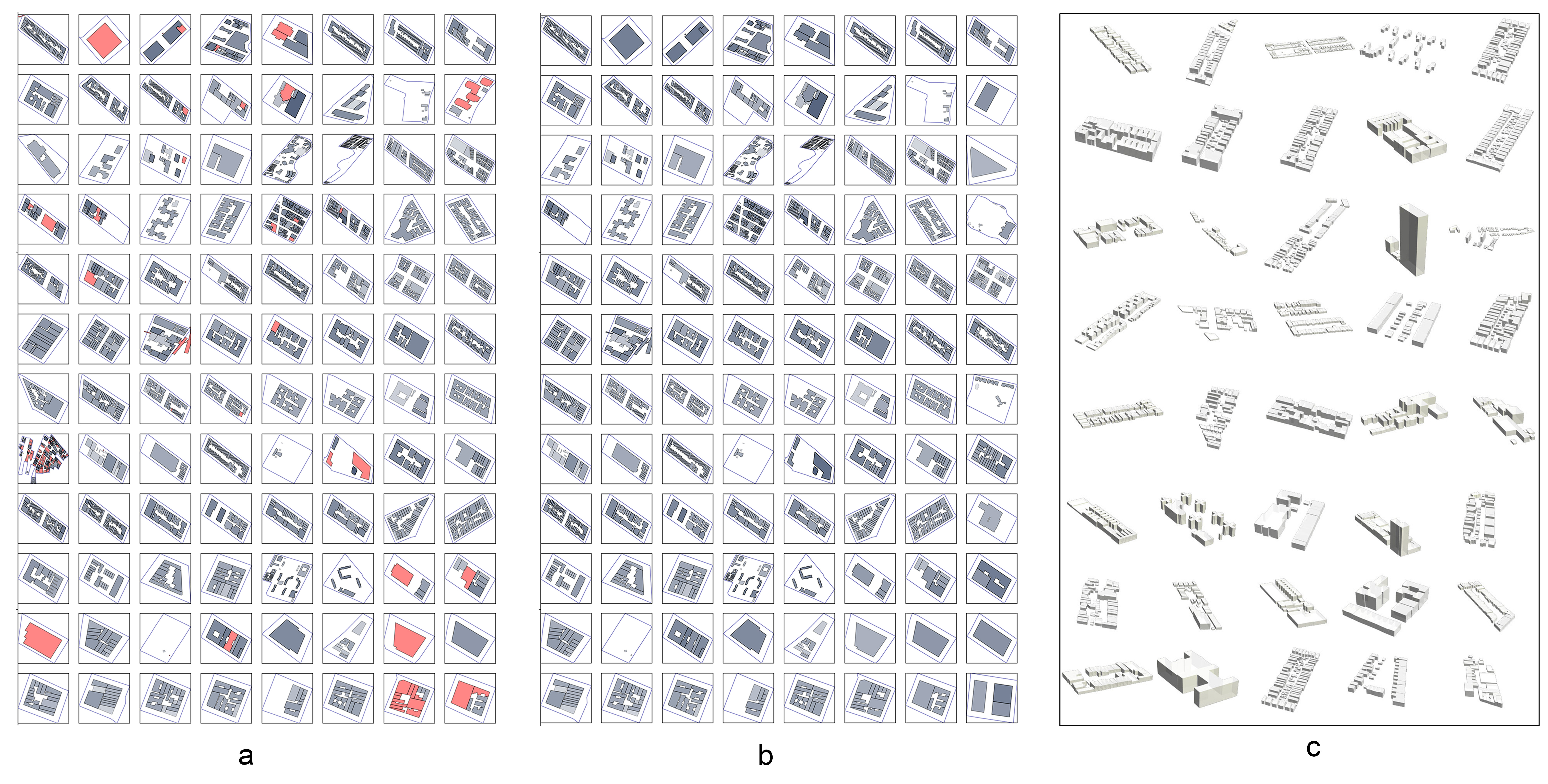}
\centering
\caption{The visualisation for a) original blocks, b) after pre-processing including adding height value and filtering out the blocks still missing height information and c) samples for block-scale 3D models. The buildings that missing height information are coloured in red.}
\label{fig:heightassign}
\end{figure}

When building footprints in the dataset were missing corresponding height information (e.g. building footprints marked red in Figure \ref{fig:heightassign}a ), we assigned the average height of the surrounding buildings. While this average value likely does not correspond to the actual height, we argue this fuzzy estimation is sufficiently accurate to demonstrate our general methodology. This resulted in a dataset of 14,248 3D city blocks. Figure \ref{fig:heightassign}b shows that the height information is completed as no building or block is coloured red. Some of the selected 3D city blocks can be visualized in Figure \ref{fig:heightassign}c. 
In our dataset, urban blocks are defined based on the road network categories from OpenStreetMap (OSM), including Primary, Secondary, and Residential roads. 
As a result, the block scale varies significantly, covering both small urban blocks and large superblocks (combination of multiple mini-blocks). 
The dataset includes blocks of diverse scales, with the smallest blocks measuring just over 6,000 m² and the largest exceeding 980,000 m².
The block scale variations make the dataset representative of various urban form. 
The variation is meaningful for testing the robustness of our methodology, as it ensures applicability across various block-scale urban form.

\subsection{Urban morphology indicator selection}

Regarding urban form indicators, quantitative studies have classified urban morphology indicators (UMIs) into six categories: dimension, shape, spatial distribution, intensity, connectivity, and diversity \citep{fleischmann2021measuring}.
Among these, dimension (i.e., height), shape, and intensity can be directly measured at the individual building level, whereas spatial distribution, connectivity, and diversity require quantification at larger scales, such as streets or larger built-up areas.
Given our focus on block-scale building morphology, we have selectively chosen metrics related to dimension, shape, and intensity. 
Additionally, we incorporate performance-related indicators (e.g., meteorological modelling indicators) due to their effectiveness in both evaluating local climate and characterising outdoor spaces within 3D city blocks \citep{burian2002morphologicalanalysis}.
These indicators also provide a more comprehensive representation of height variations and surface ratio quantifications, enabling the selection of UMIs that can be applied to further enhance 3D block characterisation and integrate urban performance evaluation.

We identified 15 morphological indicators that are globally used for urban planning practice and urban performance evaluation.
They represent 4 types of block characteristics: height, surface ratio, block content and block 2D shape. Table \ref{table:UMIs} provides definitions for each indicator, detailing their relevance in representing the urban form characteristics of a block.
The building coverage ratio (BCR), also referred to as site coverage and the ground space index (GSI), and the floor area ratio (FAR), also known as the floor space index (FSI), 
These indicators are among the widely recognised metrics in both urban form studies \citep{berghauserpontSpacemateDensityTypomorphology2005a, pan2008analyzingvariationofbuildingdensity, soliman2018quantifyingBCR,li2024characterizing} and urban performance evaluations, such as microclimate on street canyons and energy consumptions \citep{jung2021analysis,cheshmehzangi2021towards}.
Various height-related metrics are considered, including minimum height (MinH), average height (AveH), standard deviation of height (SDH), and weighted average height (WAH), to comprehensively capture the vertical characteristics of a group of buildings.
Meanwhile, they serve as key indicators for urban environmental assessment, such as for evaluating urban heat islands and ventilation in canyons street canyons \citep{hang2012influenceofbuildingheight,karimimoshaver2021effectofUM,burian2002morphologicalanalysis}.
Additionally, we incorporate morphological indicators used in urban meteorological models and diffusion models, such as complete aspect ratio (CAR) and global height-width ratio (GHWR). 
CAR and GHWR are selected for their geometric significance in capturing building texture within blocks and in representing the degree of aggregation or dispersion of building volumes, providing comprehensive representations for building morphology \citep{burian2002morphologicalanalysis}.
To represent 2D block shape features, indicators such as Block shape factor (BSF) and block shape squareness (BSS) are utilized. Additionally, average stories (AS) and open space ratio (OSR) were introduced by \citep{berghauserpontSpacemateDensityTypomorphology2005a}.

\begin{footnotesize}
\begin{longtable}{p{22mm} p{35mm} p{50mm} p{30mm}}
\caption{Overview of our selected urban morphological indicators, including their definitions and the urban form characteristic types they describe. The table shows an example of a 3D model and its indicator values.}
\label{table:UMIs} \\
\toprule
Characteristic type & UMI & Definition & Example   \\
\midrule
\endfirsthead

\toprule
Characteristic type & UMI & Definition & Example \\
\midrule
\endhead

\midrule
\endfoot

\bottomrule
\endlastfoot

\multicolumn{3}{p{90mm}}{} & \includegraphics[height=30mm]{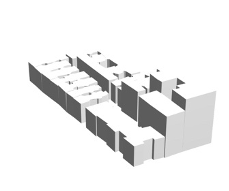} \\

\midrule
Height & MaxHeight \newline (MaxH) /m 
& Maximum building height within a block. & \centering 50.4 \tabularnewline

\midrule
Height & MinHeight \newline (MinH) /m 
& Minimum building height within a block. & \centering 18.1 \tabularnewline

\midrule
Height & AverageHeight \newline (AveH) /m 
& Average building height within a block. & \centering 23 \tabularnewline

\midrule
Height & Standard-\newline DeviationHeight \newline (SDH) 
& The standard deviation of building heights within a block.
\newline \(\displaystyle sdh = \sqrt{\frac{\sum_{i=1}^{N}{(h_{i}-\bar{h})}^{2}}{N-1}} \) \newline
$\bar{h}$ is the average height of buildings within the block. $h_{i}$ is the height of building $i$.  & \centering 9.18 \tabularnewline

\midrule
Height & Weighted-\newline AverageHeight \newline (WAH) /m 
& Building height is weighted by the building footprint area.  
\newline \(\displaystyle wah = \frac{\sum_{i=1}^{N}A_{i}h_{i}}{\sum_{i=1}^{N}A_{i}} \) \newline
$h_{i}$ is the height of building $i$. & \centering 26.15 \tabularnewline
\\

\midrule
Height &  AverageStoreys (AS) 
& Average buildings stories within a block.  & \centering 7.88 \tabularnewline

\midrule
SurfaceRatio & Building-\newline CoverageRatio (BCR) 
& The ratio of the total building footprint area to the gross block area.  
\newline  \(\displaystyle bcr = \frac{\sum_{i=1}^{N}A_{i}}{A_{b}} \) \newline
$A_{i}$ is footprint area of building $i$. $A_{b}$ is gross block area. $N$ is number of buildings in the block. & \centering 0.41 \tabularnewline

\midrule
SurfaceRatio & FloorAreaRatio (FAR) 
& The ratio of a building's total floor area to the gross area of its block.  
\newline \(\displaystyle far = \frac{\sum_{i=1}^{N}\sum_{j=1}^{S}A_{ij}}{A_{b}} \) \newline
$A_{ij}$ is the area of floor $j$ of building $i$. $S$ is the number of stories of building $i$.  & \centering 2.67 \tabularnewline

\midrule
SurfaceRatio &  CompleteAspectRatio (CAR) 
& Quantifies building envelope exposure per block using the formula 
\newline \(\displaystyle car = \frac{A_{W} + A_{R} + A_{G}}{A_{b}} \) \newline
$A_{W}$ is the total area of building vertical surfaces. $A_{R}$ is the rooftop area. $A_{G}$ represents exposed ground area within the block. & \centering 3.68 \tabularnewline

\midrule
SurfaceRatio & OpenSpaceRatio (OSR) 
& Measures the proportion of a block’s area open to the sky, calculated as 
\newline \(\displaystyle osr = \frac{A_{b}-\sum_{i=1}^{N}A_{i}}{\sum_{i=1}^{N}A_{i}L_{i}} \), \newline
where $A_{b}$ is the block footprint area. $A_{i}$ is the ground floor area of building $i$. $L_{i}$ is the number of storeys. & \centering 0.22 \tabularnewline

\midrule
BlockContent & GlobalHeight-\newline WidthRatio (GHWR) 
& Estimates the building height-to-width ratio within a block, calculated as
\newline \(\displaystyle ghwr \cong \frac{\bar{h}}{\bar{W}} \quad \bar{W} = \frac{L_{d}}{N-1} \) \newline
$\bar{W}$ is the average distance between buildings according to Delaunay algrithm. $L_{d}$ is the total length of the Delaunay network. $N$ is the number of buildings in the block. &  \centering 0.68 \tabularnewline

\midrule
BlockContent & NumberOfBuildings (NOB) 
& Total number of buildings within a block. & \centering 16 \tabularnewline

\midrule
Block2DShape & BlockArea (BA) /m\textsuperscript{2} 
& Gross area of a block footprint. &  \centering 15,718 \tabularnewline

\midrule
Block2DShape &  BlockShapeFactor (BSF) 
& Gross area of block footprint divided by the corresponding bounding box aligned with the coordinate axes. & \centering 0.47 \tabularnewline

\midrule
Block2DShape & BlockSquareness (BSS) 
& Gross area of block footprint divided by the corresponding minimum bounding box. & \centering 0.89 \tabularnewline

\end{longtable}
\end{footnotesize}

\begin{figure}[h!]
\includegraphics[width=\textwidth]{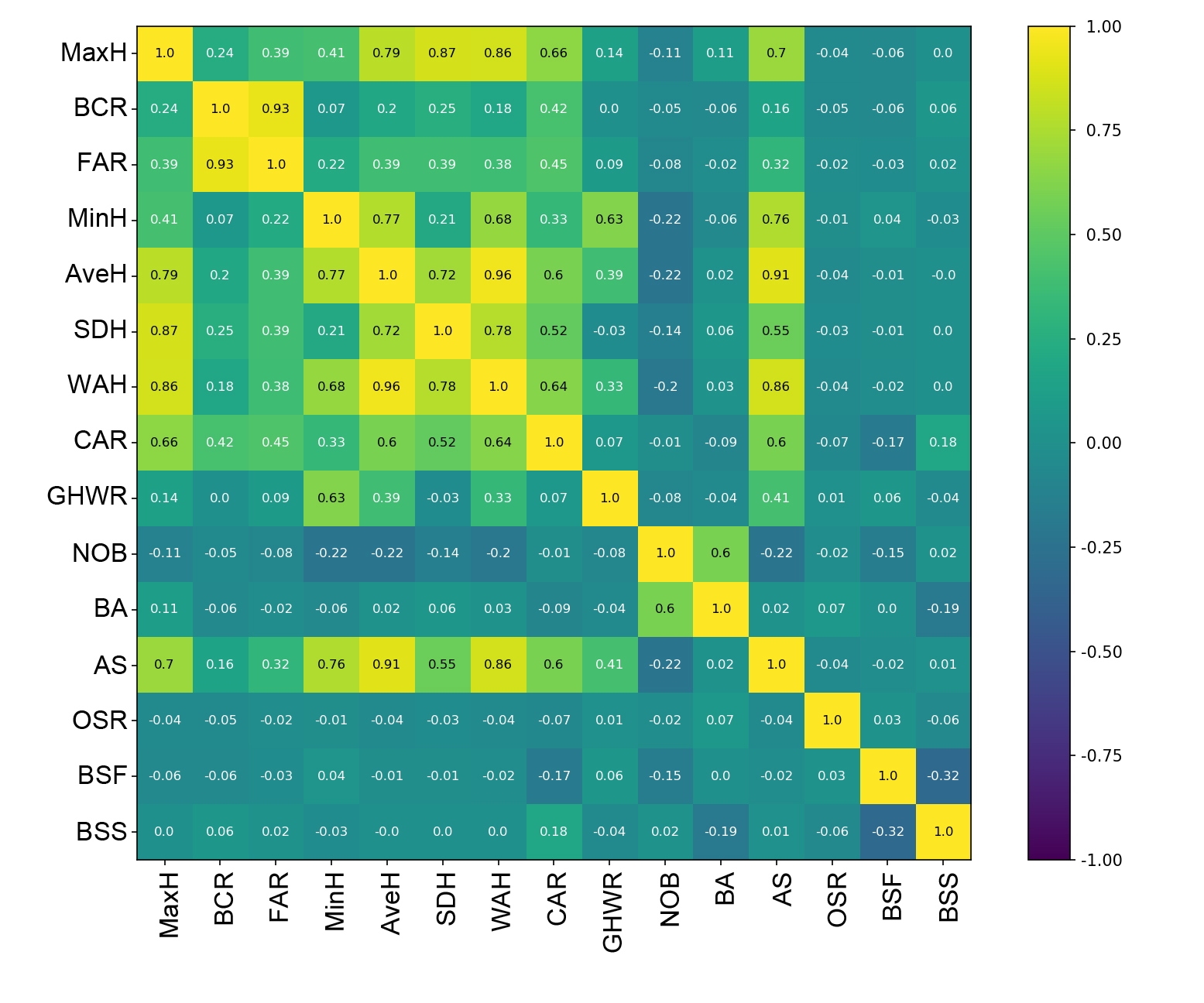}
\centering
\caption{Pearson correlation heatmap of 15 morphological indicators, illustrating their pairwise relationships in characterising urban form.}
\label{fig:pearson}
\end{figure}

Morphology metric sets have combinatorial effects in characterising urban form. To capture these effects, we need to identify metric sets comprising unique combinations of indicators that represent various urban form characteristics.
Morphological indicators may have different or overlapping capacities in characterising urban form. The selection of an appropriate metric set depends on their interrelationships in urban form characterisation.
To quantify the strength of pairwise linear relationships between indicators, we use the Pearson correlation coefficient \citep{cohen2009pearson}. A value of 1 indicates a perfect positive correlation, -1 represents a perfect negative correlation, and 0 signifies no linear relationship. 
Figure \ref{fig:pearson} presents a heatmap visualisation of the Pearson pairwise correlation coefficients among the 15 indicators.
Height-related indicators show more than moderate positive correlations with one another, with most of the Pearson correlation coefficients (PCC) exceeding 0.5. 
Among them, AverageHeight has a very strong positive correlation with WeightedAverageHeight (PCC = 0.96) and AverageStorey (PCC = 0.91), while WeightedAverageHeight has a PCC of 0.86 with AverageStorey. 
Among surface-ratio-related indicators, BuildingCoverageRatio has a strong positive relationship with FloorAreaRatio (PCC is 0.93), while less than moderate correlations appear among the other indicators. 
FloorAreaRatio has generally stronger positive correlations (higher total PCC) with the other 14 indicators.

When developing morphology metric sets, we established AllBlockMetric as a comprehensive set that includes all the indicators to cover more comprehensive characteristics (see Table \ref{table:morphosets}). 
AllBlockMetric was initially designed to facilitate effective morphology analysis.
Training a Self-Organising Map (SOM) using all 15 indicators requires longer computation time, which may be a concern in application scenarios involving large-scale optimisation, where minimising computational time at each iteration is essential.
Therefore, we developed a set that includes one block metric per characteristic type (OneBMC), a more streamlined metric set that selects one key indicator from each urban form characteristic type, to reduce computational cost when it is applied in large-scale optimisation models. 
The indicator with the strongest correlation within its category was chosen to maintain representativeness.
Therefore, WeightedAverageHeight, BuildingCoverageRatio, and BlockArea were selected.
Since NumberOfBuilding (NOB) has a negative correlation with GlobalHeightWidthRatio within the BlockContent category, but has a weaker correlation with other indicators in the set, we included NOB to enhance metric diversity. As a result, OneBMC set consists of WeightedAverageHeight, BuildingCoverageRatio, NumberOfBuilding, and BlockArea (Table \ref{table:morphosets}).

In addition to the proposed metric sets, we include two comparative sets derived from existing research -- Spacemate \citep{berghauserpontSpacemateDensityTypomorphology2005a} and BlockShape \citep{xuRenewalDesignofRoma2019}.
Spacemate is widely recognised in academic and research fields, particularly in urban form classification, density analysis, and typology comparisons \citep{steadman2014density}. It provides a structured approach to understanding and classifying urban form based on density-related indicators which have high correlations with each other, bridging the gap between quantitative and qualitative urban analysis. Our analysis shows that Spacemate’s UMIs primarily capture surface ratio characteristics, yet its effectiveness has not been evaluated through computational methods.
The BlockShape metric set was developed to integrate digital methods into the urban design of a disused area around Roma Termini.
Block shape is closely related to the urban form in Roma.
Hence, by referring to blocks with similar shapes within the development area based on morphological indicators, it aimed at informing urban design proposals and ensuring alignment with the existing city context.

We identified four distinct morphology metric sets, each comprising a unique combination of indicators that represent various urban form characteristic types.
The morphology metrics compositions are shown in Table \ref{table:morphosets}.
We propose AllBlockMetric including 15 UMIs. This comprehensive set is designed for higher precision and consistency in urban form characterisation and retrieval.
Additionally, we introduce OneBMC, a streamlined set comprising four representative UMIs from each urban form characteristic type. This set is designed for lower computational cost when it is applied for large-scale urban performance optimisations.
It is potentially suitable for urban studies involving large datasets where a lower level of precision and consistency in urban form representation is acceptable.
To compare with our proposed metric sets, we incorporate two existing sets from the literature --- one focusing on density-related indicators and the other on 2D shape-related indicators --- as they provide valuable insights for their respective aspects of urban form classification and digital urban design applications.

\begin{footnotesize}
\begin{longtable}{p{30mm} p{18mm} p{18mm} p{20mm} p{20mm} p{22mm}}
\caption{Overview of the identified urban morphology metric sets and their compositions. The table shows that different morphology metric sets include indicators spanning various characteristic types. We propose one set that includes all 15 indicators (AllBlockMetric) and one set that contains one block metric per characteristic type (OneBMC)} \label{table:morphosets} \\

\toprule
Morphology metric sets & Height & SurfaceRatio & BlockContent & Block2DShape & Resource\\
\midrule
\endfirsthead

\multicolumn{5}{c}{\textit{(Continued from previous page)}} \\
\toprule
Morphology metric sets &  Height & SurfaceRatio & BlockContent & Block2DShape & Resource \\
\midrule
\endhead

\midrule
\multicolumn{5}{r}{\textit{(Continued on next page)}} \\
\endfoot

\bottomrule
\endlastfoot

Spacemate & AveS & BCR; \newline FAR; \newline OSR & ~ & ~ & \cite{berghauserpontSpacematrixSpaceDensity2021a}\\
\midrule
BlockShape & ~ & FAR & ~ & BA; \newline BSF; \newline BSS  & \cite{xuRenewalDesignofRoma2019}\\
\midrule
OneBMC & WAH & BCR & NOB & BA & Proposed by this paper\\
\midrule
AllBlockMetric & MaxH; MinH; \newline SDH; AS; \newline AveH \newline WAH & BCR; \newline FAR; \newline CAR; \newline OSR & GHWR; NOB & BA; \newline BSF; \newline BSS & Proposed by this paper \\

\end{longtable}
\end{footnotesize}

\subsection{Urban form clustering based on morphology metric sets}
\label{sec:somresults}

We deployed clustering techniques using the self-organising map (SOM). 
By comparing the clustering results based on different morphology metric sets, we can validate their performance. 
A Self-Organising Map (SOM), also known as a Self-Organising Feature Map (SOFM), is an unsupervised machine learning algorithm used for dimensionality reduction and data visualisation \citep{kohonen1982self, kohonen2013essentials}. 
It maps high-dimensional data onto a lower-dimensional (typically two-dimensional) grid while preserving the topological structure of the data. 
SOMs use a competitive learning approach where neurons in the network adjust their weights based on input data, grouping similar data points closer together. 
SOMs offer a unique approach to structural learning by extracting relationships directly from observed data, rather than imposing a predefined function. Unlike classical clustering modelling, which fits data to a fixed structure by minimising error deviations, SOMs preserve the underlying logic of the data, clustering similar instances while maintaining their individuality. This makes SOMs particularly effective for nonlinear function approximation and pattern mining \citep{vesanto2000clustering, moosavi2015computational,cai2022dataclusteringinUCM}. 
Clustering while keeping the topological structure of the data makes SOMs useful for clustering and exploratory data analysis in various domains. Hence, SOMs can facilitate both clustering and flexible case retrieval. 

Two simultaneous processes explain the SOM algorithm (formula \ref{eq:somformula1}, \ref{eq:somformula2}). The training data set can be considered as $ X $ = $ x_{1} $, $...$, $ x_{N} $ , as a set of N points in an n-dimensional space. A SOM can be considered as a grid with $ K $ nodes, with a set of indices $ y_{j} $, each attached with a high-dimensional weight vector, $ w_{j} $ . During the training process, an index $ y_{j} $ can be assigned for each data $ x_{i} $. The index is also called the best-matching unit (BMU). $s(w_{j} , x )$ is a similarity function which is calculated by the inverse of the distance between the input sample feature vectors $x$ and the weight vectors of the SOM node $j$, $w_{j}$.

\begin{equation}
BMU(x)=\max x_{j}\left ( f(y_{j}|x) \right )
\label{eq:somformula1}
\end{equation}

\begin{equation}
f(y_{j}|x)=\frac{\exp s\left ( w_{j},x\right )}{\int_{j=1}^{k} \exp s\left ( w_{j},x\right)}
\label{eq:somformula2}
\end{equation}

We trained four self-organising maps (SOMs) for clustering, using normalised morphology metric values as urban form feature vectors for input. 
In our experiments, the four SOMs were structured on a 10 × 10 grid. 
The neuron weights of each SOM were randomly initialised at the start point for training, with the same dimensions as the number of UMIs (e.g., 15 dimensions for the AllBlockMetric set).
During the SOM training, every sample was used for updating neuron weights per iteration.
Euclidean distance was used as the distance function, which is standard for SOMs. 
For each sample during one iteration, the best match unit (BMU) --- the nearest neuron to the input sample --- was found based on the distance function.
We used the Gaussian function as the neighbourhood function, which was used to control how much influence a weight vector has on the update during training.
The model setup was that the learning rate and the neighbourhood influence radius decrease as the number of iterations increases. We trained the SOMs with 1000 iterations.
The neuron weights were updated gradually to be closer to the input samples so that the map was self-organised and captured the dataset's structure.
After training, the input samples were clustered and indexed according to the BMU they fall within. 
In other words, each SOM neuron was associated with a set of 3D urban form models based on distance.

\begin{figure}[h!]
\includegraphics[width=\textwidth]{images/ColoredSOMs-OneBMC-AllBlockMetric.jpg}
\centering
\caption{The clustering results for the OneBMC set (left) and AllBlockMetric (right) using self-organising maps. The neurons are colour-coded based on their weights, and the 3D model visualised in each neuron is randomly selected from the models for which that neuron serves as the best matching unit (BMU).}
\label{fig:SOMs}
\end{figure}

A common way of visualising the final output of a SOM is to visualise one of the input data that is associated with the neurons since a set of input data is assigned to each neuron. 
Figure \ref{fig:SOMs} is the visualisation of two SOMs that are trained with the OneBMC set and AllBlockMetric set, respectively. 
Each neuron is colour-coded by mapping its high-dimensional weight value to a three-dimensional RGB value for visualisation. 
The 3D model visualised in each neuron is randomly selected from the samples for which that neuron serves as the best matching unit (BMU).
The colour map visualises the gradients of the trained SOM, but does not necessarily demonstrate the neuron weights.

The colour gradient indicates whether the trained SOM is smoothly distributed across the data space, without abrupt transitions.
If neighbour neurons exhibit similar colours, it suggests that they have similar weight values in certain dimensions. 
Consequently, if the morphology metrics effectively characterise the 3D models, the corresponding 3D models should also share certain similarities.
Thus, if the spatial correlations between 3D models in neighbouring neurons align with the colour-based correlations of the neurons, it indicates that the morphology metrics effectively capture the 3D model characteristics.
From initial observations, both SOM maps display colour gradients, indicating that the networks are trained to smoothly capture the structure of the data space. 
The visualised 3D models vary across different neurons and also exhibit gradual transitions in urban form characteristics. 
AllBlockMetric SOM appears to have a smoother transition of 3D models than OneBMC SOM. 
For example, in the blue region of OneBMC, a mix of triangular and rectangular block shapes appears. This variation occurs because OneBMC lacks the block shape-related indicator.
In AllBlockMetric SOM, neighbouring neurons with similar colours (e.g., the top-right green area) exhibit 3D models with certain similar characteristics too.
Overall, the maps suggest that AllBlockMetric captures urban form characteristics more comprehensively, while OneBMC provides a more condensed representation with certain limitations.

\begin{figure}[h!]
\includegraphics[width=\textwidth]{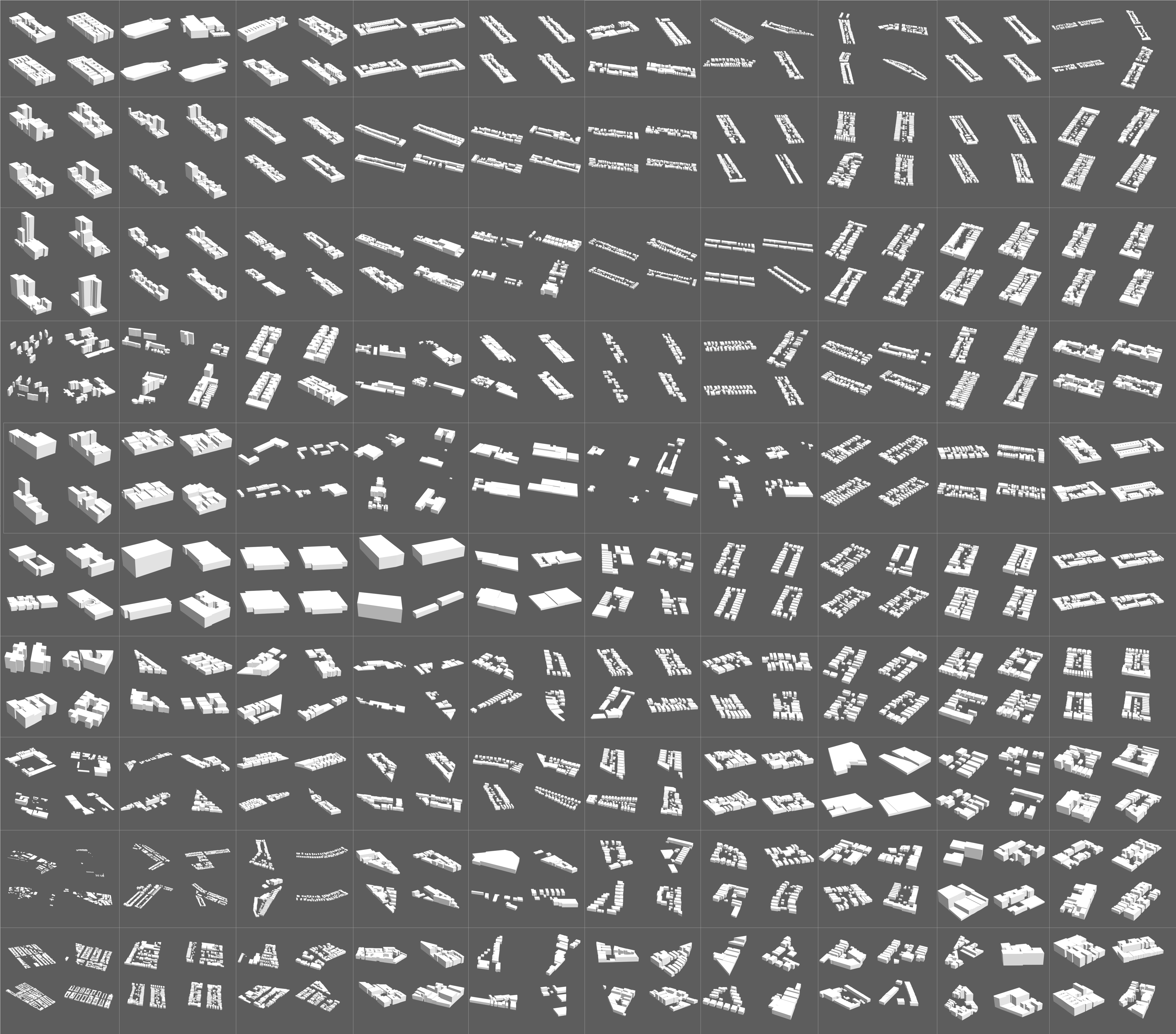}
\centering
\caption{The clustering results for AllBlockMetric using self-organising maps, visualising four 3D models in each neuron as their BMU. It shows clear clustering as well as gradual transitions between clusters }
\label{fig:allmetricSOM}
\end{figure}

In the AllBlockMetric SOM map, we visualised 4 models in each neuron for further assessment (Figure \ref{fig:allmetricSOM}).
Regarding 3D models, we observe clear clustering as well as gradual transitions between clusters.
For example, we can identify clusters of blocks with a single large building (middle-left), high-density compact buildings (top-left), and linear-shaped blocks with inner spaces and regularly spaced surrounding buildings (top-right), and so on.
The transitions between clusters are also exhibited. From top-left to top-right, there is a progressive shift from compact, high-density buildings to blocks with a looser building texture. Similarly, from bottom-left to bottom-right, we observe a transition from low-density superblocks to mid-density triangular blocks, to higher-density compact blocks.
The other three SOM maps can be found from\ref{sec:soms:appendix}. 
More detailed comparisons and similarity analysis based on the clustering results will be introduced in the later sections.

\subsection{Context-based case retrieval using SOM encoding}

\begin{figure}[h!]
\includegraphics[width=\textwidth]{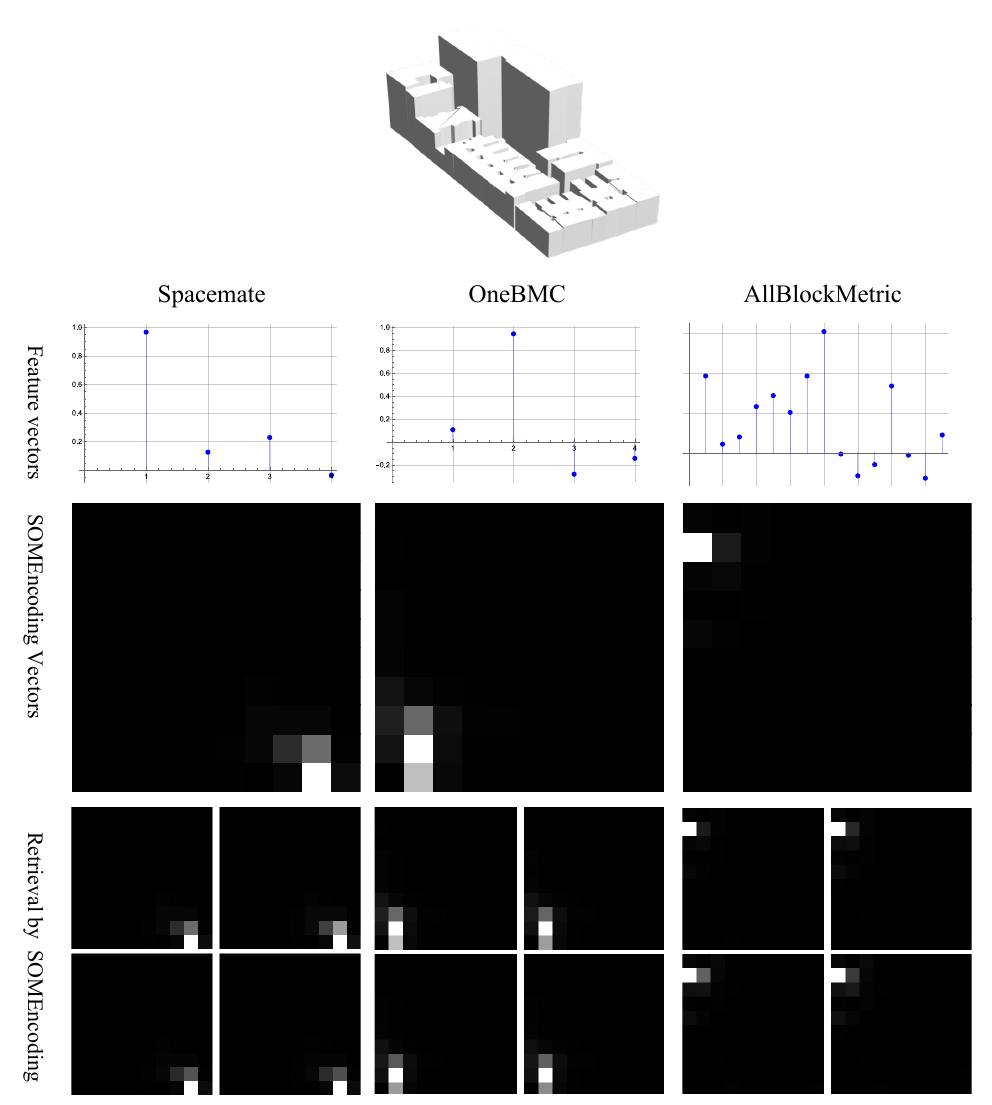}
\centering
\caption{The figure showcases the feature vectors of three morphology metric sets and visualises the SOMEncoding value based on the morphology metrics. The retrieved cases based on SOMEncoding vectors show high similarity.}
\label{fig:somebedding}
\end{figure}

Each 3D model sample's characteristic is encoded by feature vectors which are normalised morphology metrics. 
Each sample has four sets of feature vectors with varying dimensions, including Spacemate, BlockShape and OneBMC feature vectors with 4 dimensions, and AllBlockMetric feature vectors with 15 dimensions.
For instance, Figure \ref{fig:somebedding}, shows three kinds of normalised indicators based on Spacemate, OneBMC and AllBlockMetric, taking one sample as an example.
The similarity of the 3D models is determined by the distance between their feature vectors. One of the common ways is to calculate the Euclidean distance based on certain feature vectors. Hence similar cases can be retrieved and ranked based on their distance. During the case retrieval process, the calculated distance based on feature presentation significantly influences the retrieval results.

In this study, SOMs serve two primary roles --- clustering and encoding for context-based case retrieval. 
To facilitate context-based retrieval, we developed SOMEncoding vectors, which are used to measure similarity based on Euclidean Distance (ED) rather than relying on the Best Matching Unit (BMU) or directly using normalised original indicators.
While BMU-based case retrieval is a common approach in SOM applications, it has limitations. For instance, if a SOM is highly curved in certain regions, two data points that are very close in feature space may be assigned to two different BMUs, leading to inconsistencies in retrieval.
Alternatively, case retrieval using ED between feature vectors is straightforward and computationally efficient, particularly in low-dimensional spaces (e.g., 4D). 
However, in higher-dimensional spaces (even in 15D), data points tend to become nearly equidistant, reducing the discriminative power of Euclidean distance when applied directly to the original indicators.
To address this, context-based encoding can improve the feature representation \citep{zhang2018context}.
A trained SOM provides a structured representation of the data space, capturing its underlying patterns. By further encoding the normalised original indicators using a trained SOM, we can have a more context-aware representation, leading to enhanced retrieval performance.

In our SOMEncoding method, vectors are calculated through the distance between normalised morphology metrics and the corresponding trained SOM's neuron weights. The feature vectors computed based on SOMEncoding will be referred to as SOMEncoding vectors.
By applying a context-based encoding method, it captures the unique characteristics of a sample relative to the entire data space. 
For a sample whose feature vector (normalised morphology metric value) is $ x $. If the SOM map has $K$ neurons, the SOMEncoding vector is defined as the set of Euclidean distances between $ x $ and every SOM neuron weight, $ w_{j} $, where $j = (1,..,K)$. Hence the SOMEncoding vector $ s $ has $K$ dimensions (in our case, 100 dimensions). The SOMEncoding vector $ s $ for the sample $ x $ can be described as Formula \ref{eq:somencoding}, where the Euclidean distance between \( x \) and \( w_j \) is computed as Formula \ref{eq:somencodingdistance}, where \( x_{m} \) is the \( m \)-th dimension feature of the sample \( x_i \), \( w_{jm} \) is the \( m \)-th weight component of the neuron \( j \) in the SOM.

\begin{equation}
s = \left[ d(x, w_1), d(x, w_2), \dots, d(x, w_K) \right]
\label{eq:somencoding}
\end{equation}

\begin{equation}
d(x, w_j) = \sqrt{\sum_{m=1}^{M} (x_{m} - w_{jm})^2}
\label{eq:somencodingdistance}
\end{equation}

The morphology metrics were re-encoded into 100-dimensional SOMEncoding vectors. We visualised the SOMEncoding vectors by grayscale images in a 10x10 map, taking a selected block sample as an example, showcasing Spacemate, OneBMC and AllBlockMetric  (Figure \ref{fig:somebedding}). 
The closer the sampled vector is to a SOM neuron, the lighter its corresponding colour appears.
For example, in Figure \ref{fig:somebedding}, the sample's Spacemate vector is close to three SOM neurons, with the closest neuron (white-coloured) located at the bottom-right, which will be identified as the Best Matching Unit (BMU) of the sample. A similar pattern --- the sample vector is close to 3 SOM neurons --- is observed for the OneBMC vector.
This demonstrates that when case retrieval is based solely on BMUs, discrepancies can arise because cases with different second-closest or third-closest neurons can be identified as well. 
In contrast, retrieving cases using the SOMEncoding vector improves precision, as it captures the full distance-based map rather than relying only on the BMU.
Figure \ref{fig:somebedding} also showcases the four grey-scale colour maps which are the top four closest cases according to SOMEncoding vectors. 
The retrieved four grey-scale colour maps show a high degree of consistency with the targets.

SOMEncoding vector offers the advantage of making case retrieval across different morphology metric sets more comparable and improves the precision. By reducing the heterogeneity between the morphology metrics, all samples can be represented as vectors in the same dimensions (100 dimensions) after encoding. To summarise the whole encoding process, we mapped a 3D urban form to its corresponding morphology metrics, then normalised the morphology metrics into feature vectors based on which we trained SOMs, and then encoded the feature vectors based on the context of trained SOMs. Finally, we can use the morphology metrics to link more similar and diverse urban forms through case retrieval.

\section{Results}
\label{sec:result}

\subsection{Mapping between morphology metrics and 3D urban forms}

\begin{figure}[h!]
\includegraphics[width=\textwidth]{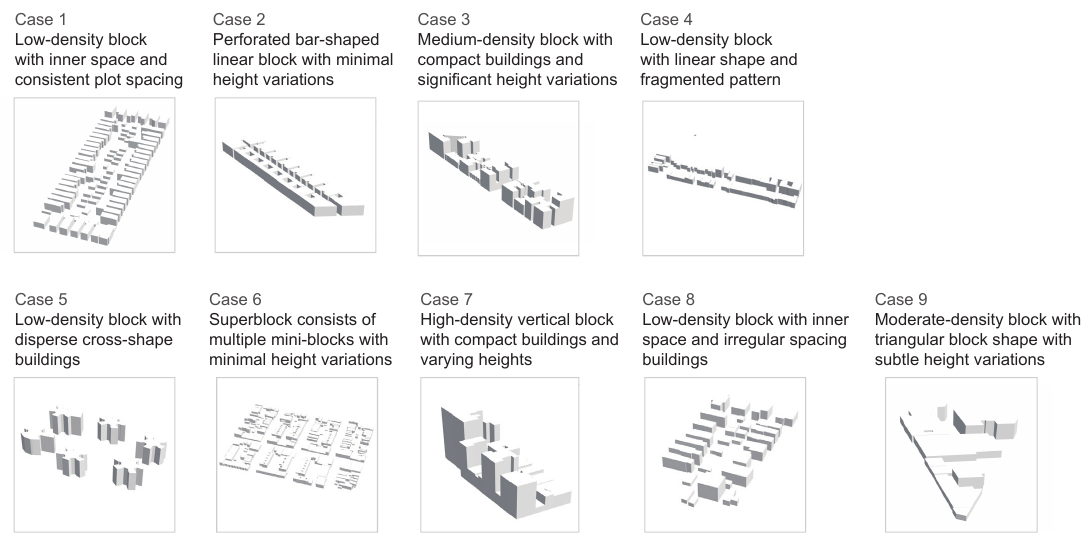}
\centering
\caption{Nine cases vary in morphological prototypes are selected for validating morphology metrics in representing 3D urban forms.}
\label{fig:cases}
\end{figure}

To assess the effectiveness of the four morphology metric sets in capturing urban form characteristics, we compare the similarities between the retrieved 3D models with the targets. 
Each 3D model is generated based on the target's morphology metrics using one of the four metric sets.
In this paper, we select nine cases that vary in morphological prototypes (Figure \ref{fig:cases}).
To evaluate the effectiveness of different morphology metric sets in representing urban form characteristics, we selected real-world cases from New York City (NYC). These cases were chosen to reflect a different range of density levels, height variations, building textures, and block shapes.
For example, Case 2 represents a linear block with minimal height variations, while Case 3 is a medium-density block with compact buildings and significant height variations. Case 9 features a triangular-shaped block.

In Figure \ref{fig:cases}, the selected cases include a low-density block with inner space and consistent plot spacing (Case 1), where low-rise buildings are arranged uniformly around a central public space. 
Case 2 represents a perforated bar-shaped linear block with minimal height variations, characterized by a long, bar-like structure with rhythmic rectangular voids, likely indicating courtyards or shared spaces.
Case 3 is a medium-density block with compact buildings and significant height variations, forming a dynamic rhythm along a linear alignment. Case 4, a low-density block with a linear shape and fragmented pattern, consists of irregularly spaced structures with relatively consistent heights. 
Case 5 is a low-density block with dispersed cross-shaped buildings, arranged in a loosely spaced grid pattern with substantial open spaces. 
Case 6, a superblock with multiple mini-blocks and minimal height variations, features a subdivision pattern resembling a courtyard or perimeter block configuration. 
Case 7, a high-density vertical block with compact buildings and varying heights, exhibits a dense cluster of tall buildings with minimal spacing, emphasizing verticality. 
Case 8, a low-density block with inner space and irregularly spaced buildings, consists of scattered structures with varying orientations, creating a fragmented pattern. 
Finally, Case 9 is a moderate-density block with a triangular shape and subtle height variations, characterized by a triangular site boundary, moderate density, and irregular building sizes. 
By selecting blocks with distinct morphological prototypes, we validate assessment of the morphology metric sets for their capacity to represent diverse 3D city blocks.

\subsection{Evaluating morphology metric sets by comparison}
\label{subsec:evaluate}
We evaluate the performance of morphology metric sets by comparing the 3D models that are retrieved according to the four morphology metrics. The more similar the retrieved 3D models are, the more effective the metric set is. 
Each of the nine selected cases has four types of metric sets and corresponding SOMEncoding vectors, based on which four lists of 3D models are retrieved. 
We showcase 3D models with the highest similarities to the target for further validation. 

\begin{figure}[h!]
\includegraphics[width=0.9\textwidth]{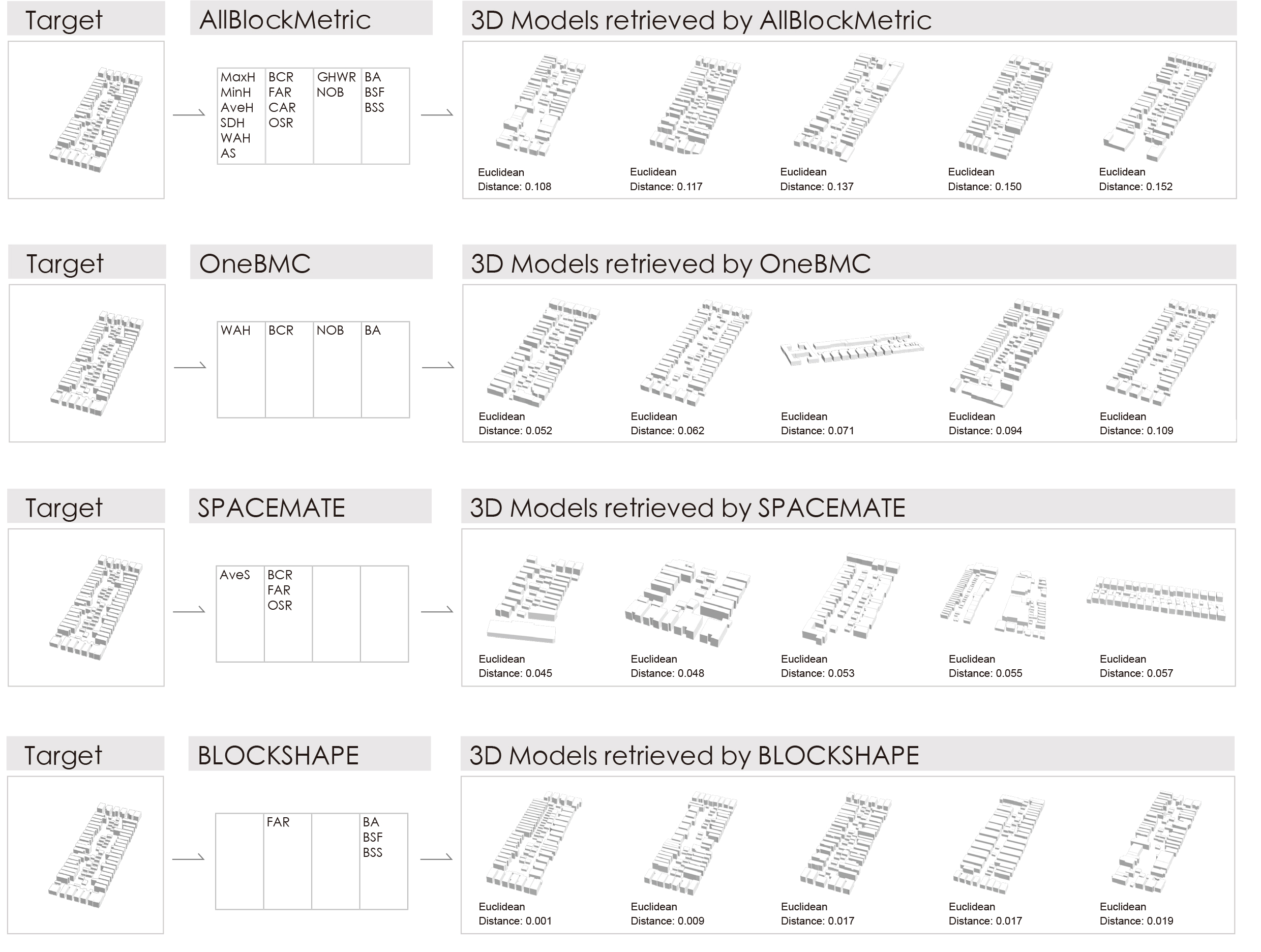}
\centering
\caption{AllBlockMetric, OneBMC, Spacemate and BlockShape are derived from the target and are used to retrieve 3D models. The figure showcases the top 5 similar 3D models. The Euclidean distance of SOMEncoding vector to the target represents their similarity; therefore, the smaller the distance, the more similar the two cases are.}
\label{fig:case1comparison}
\end{figure}

In Figure \ref{fig:case1comparison}, after setting case 1 as the target, its morphology metric sets -- AllBlockMetric, OneBMC, Spacemate and BlockShape -- are generated.
Then 3D models with the top similarities are generated according to the target's SOMEncoding vectors, respectively. 
The similarities are calculated based on the Euclidean distance to the target's SOMEncoding vector, with retrieved models ranked by their distances. In this demonstration using Case 1, 3D models retrieved via the AllBlockMetric generally exhibit very high similarities to the target case. 
The BlockShape set performed well in this case, whereas models retrieved using Spacemate exhibited significant discrepancies.
We hypothesize that in New York City, this type of urban form is more strongly correlated with block shapes rather than density-related features.
Interestingly, OneBMC achieved performance comparable to AllBlockMetric, despite the third retrieved case demonstrating lower similarity. This suggests that incorporating block content and shape factors, such as NumberOfBuildings (NOB) and BlockArea (BA), can significantly enhance urban form characterisation in this case.

\begin{figure}[h!]
\includegraphics[width=0.95\textwidth]{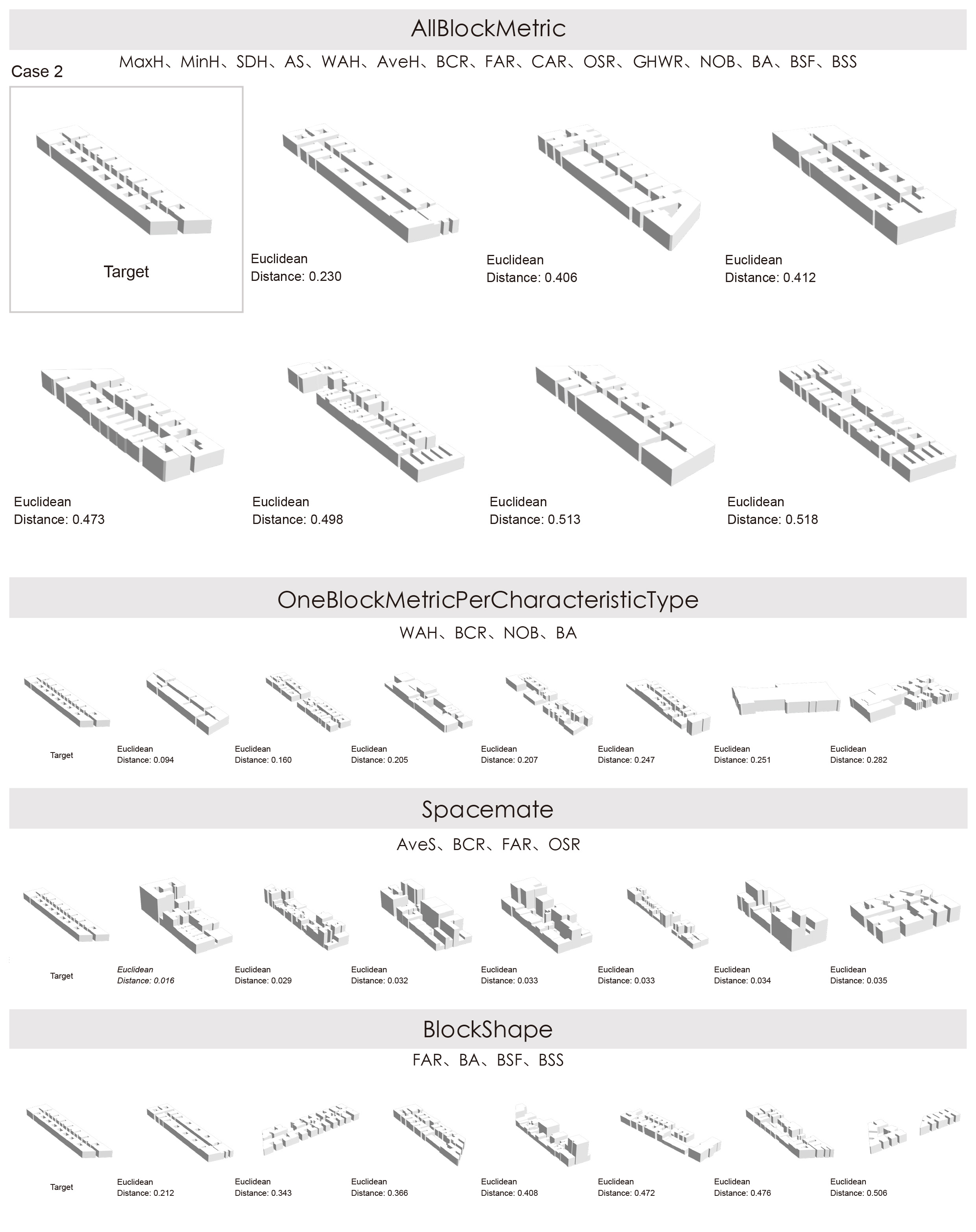}
\centering
\caption{The figure shows the results of comparisons between AllBlockMetric, OneBMC, Spacemate and BlockShape. The target is case 2 - a perforated bar-shaped linear block with minimal height variations. We compare the performance of the 4 morphology metric sets by comparing the general similarity of the retrieved 3D block-scale models.}
\label{fig:c2}
\end{figure}

In Figure \ref{fig:c2}, case 2 is set as the target. The 3D models retrieved using the AllBlockMetric exhibit a high degree of similarity to the target, all displaying structures with rhythmic rectangular voids or cutouts along their long sides. 
Cases with semi-triangular block shapes are successfully retrieved as well.
In contrast, models retrieved using Spacemate demonstrate significantly different patterns. Density-related indicators failed to capture the height variation and factors like the number of buildings in the block.
Models retrieved via the BlockShape metric retain the elongated and linear block shape but exhibit varied density and building layouts. 
Compared to BlockShape and Spacemate, the models retrieved by OneBMC exhibit greater consistency in height variations, suggesting that Weighted Average Height (WAH) effectively captures urban form height variations.

\begin{figure}[h!]
\includegraphics[width=0.95\textwidth]{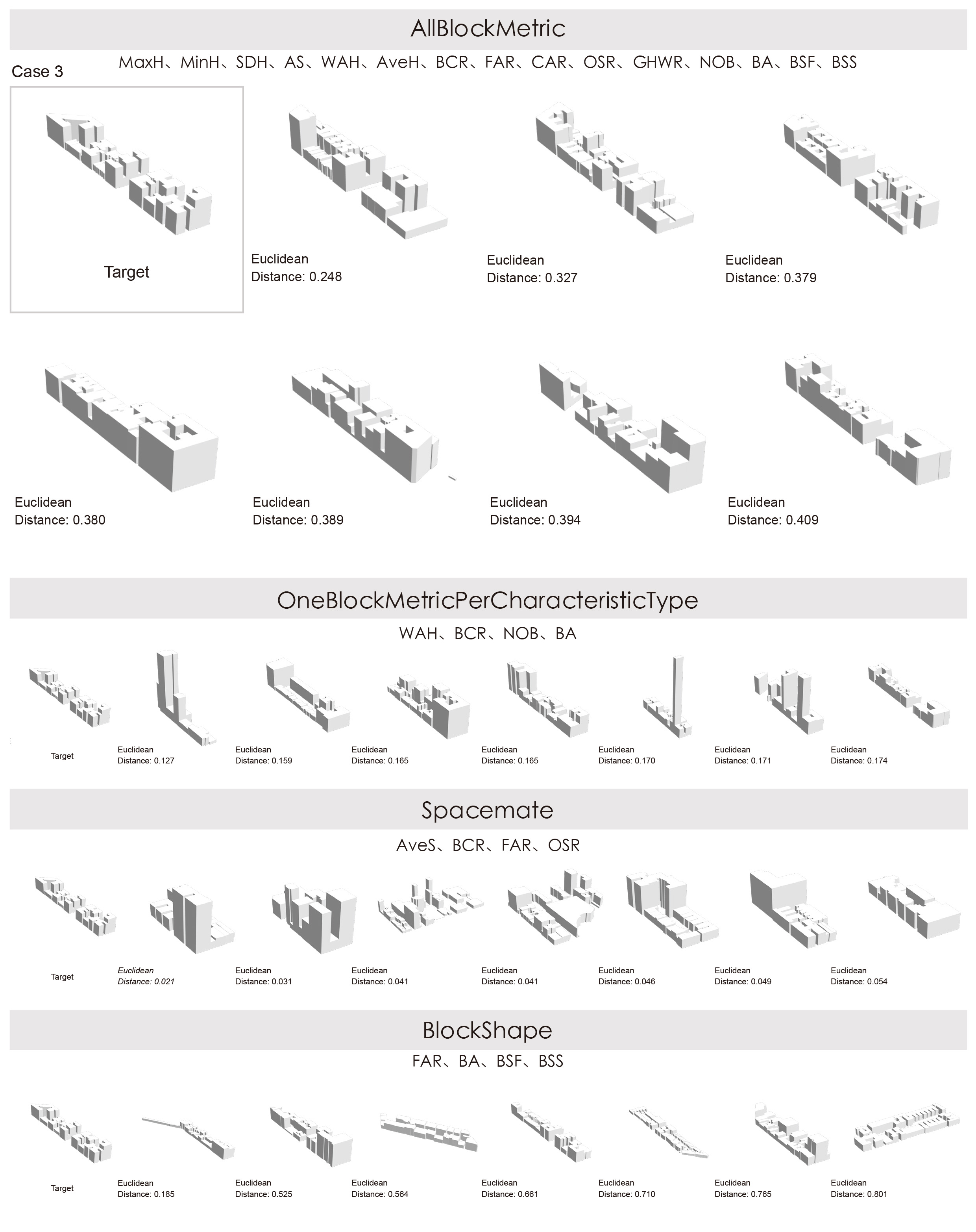}
\centering
\caption{The figure shows the results of comparisons between AllBlockMetric, OneBMC, Spacemate and BlockShape. The target is case 3 - a medium-density block with compact buildings and significant height variations. }
\label{fig:c3}
\end{figure}

\newpage
In Figure \ref{fig:c3}, case 3 is set as the target. 
Case 3 features a complex urban form characterised by varied building heights and a diverse arrangement of buildings.
Models retrieved using AllBlockMetric demonstrate a good level of similarity, capturing the medium-density block with significant height variations. 
While some variations in height and spacing, but the overall form remains consistent, especially compared with other sets.
In contrast, BlockShape failed to capture these characteristics due to the absence of height-related indicators.
Although OneBMC and Spacemate perform better than BlockShape, noticeable discrepancies in height variations and block shape remain.
Hence, in such a complex urban form, incorporating more comprehensive morphology metrics may be necessary to achieve a better characterisation.

\begin{figure}[h!]
\includegraphics[width=0.95\textwidth]{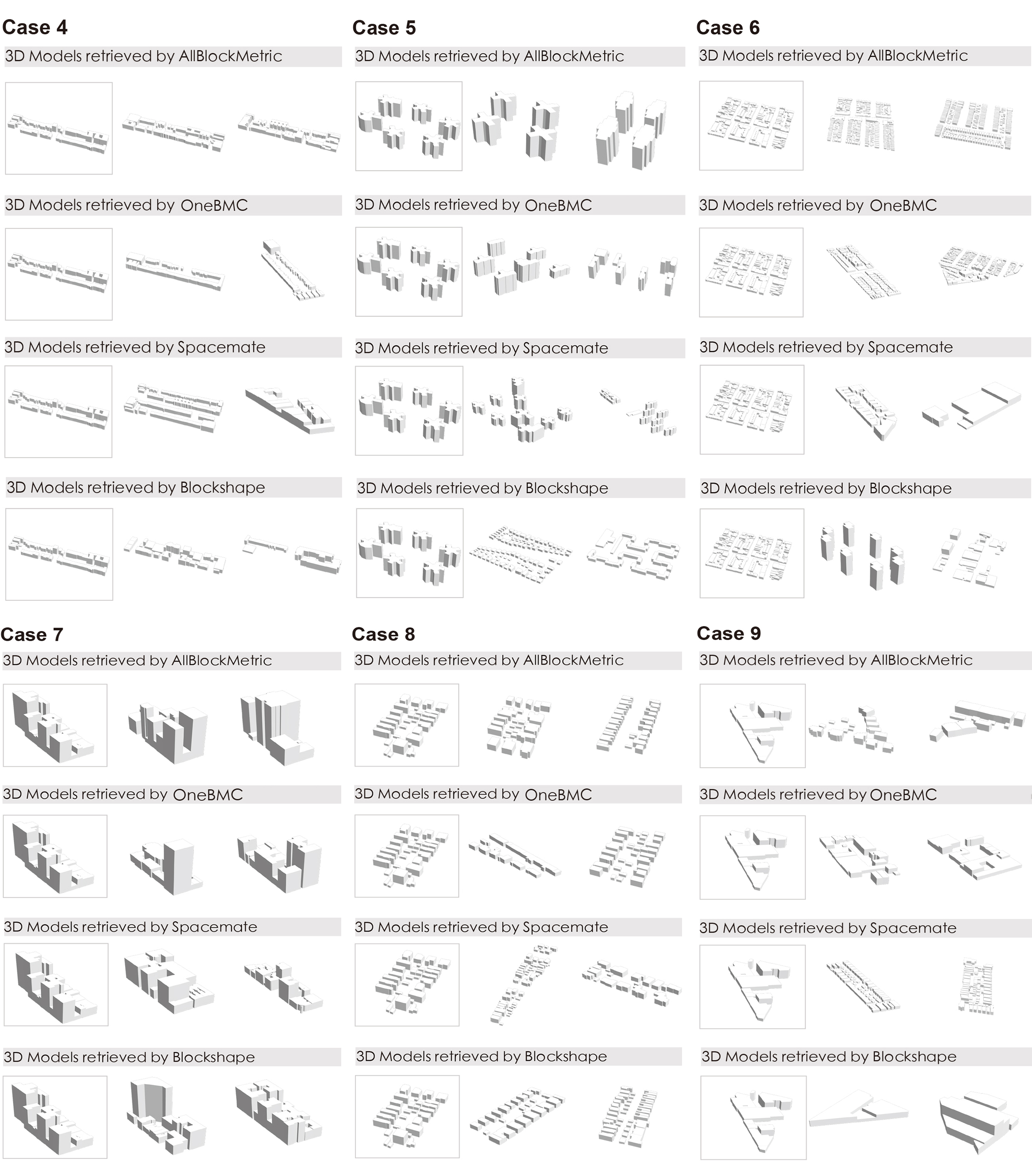}
\centering
\caption{Case retrieval results for the selected cases based on AllBlockMetric, OneBMC (one block metric per characteristic type), Spacemate and BlockShape. The 3D models in grey boxes are the targets, and the two retrieved models showcased have the highest similarity indexes. }
\label{fig:allcases}
\end{figure}

Figure \ref{fig:allcases} presents the 3D models retrieved using the four metric sets for the other six cases. 
More retrieved models are provided in \ref{sec:casestudy:appendix} for reference.
The cases exhibit different characteristics in terms of building texture, density, height variations and block shape.
The models in the grey grids are target cases.
Overall, the AllBlockMetric set effectively captures key urban form characteristics.
OneBMC, as a computationally efficient alternative, performs comparably well in some cases but exhibits noticeable discrepancies. We summarise our observation as follows:
\begin{itemize}
    \item The low-density models but with intricate building textures, such as cases 4, 5 and 6, where case 5 has cross-shaped buildings and dispersed building distributions and case 6 has multiple plots with courtyard or perimeter plot configurations. BlockShape struggled to capture the details of texture. Focusing on density indicators while lacking block content indicators, such as NumberOfBuilding (NOB), GlobalHeightWidthRatio (GHWR), Spacemate also failed to represent building texture properly. OneBMC managed to retain some density and texture characteristics but displayed notable discrepancies, such as the retrieval of a large single building (see Figure \ref{fig:c56}, case 5). It might be because NOB has inadequate influence due to the absence of CAR and GWHR. Similarly, BlockShape, which lacks sufficient density-related indicators, retrieved models with inconsistent density levels, including low-density blocks for case 3 and significant variations in case 6 (see Figure \ref{fig:c56}, case 6). It is worth mentioning that AllBlockMetric even successfully captured these intricate building textures.
    \item In cases 3 (see Figure \ref{fig:c3}) and 7, which feature complex height variations, BlockShape, Spacemate, and OneBMC --- lacking multiple height-related indicators --- failed to capture the complexity, instead retrieving models with large, monolithic buildings (see Figure \ref{fig:c78} case 7).
    \item Retrieving similar urban forms to case 8 is very challenging due to the relatively irregular pattern, characterised by scattered arrangements, varying orientations, and open spaces between buildings. 
    The BlockShape and Spacemate metrics largely fail to capture this complexity, as most of the retrieved 3D models display significantly different urban forms (also see Figure \ref{fig:c78} case 8). The 3D models retrieved via the AllBlockMetric once again demonstrate a high degree of similarity to the target, despite the challenges posed by the irregular urban form.
    \item For case 9, a triangular block,  Spacemate performed poorly in capturing the block shape. BlockShape successfully retrieved triangular blocks but with high descrepencies in building texture and density.
    OneBMC retrieved rectangular blocks, indicating some limitations in its ability to capture irregular block shapes. 
    Although AllBlockMetric retrieved models with varied urban forms (see Figure \ref{fig:c9}), it performed relatively better than other methods. This may be due to the limited number of similar city blocks in the dataset, making retrieval more challenging.   
\end{itemize}

In summary, the 3D models retrieved using AllBlockMetric demonstrate a high level of similarity to the target urban form prototypes while also capturing a diverse and rich variety of urban form.
What we established by observing the results of the case retrievals is that the UMI combinations have combinatorial effects in identifying and representing 3D urban forms. 
AllBlockMetric has good performance in comprehensively capturing the characteristics and complexity of 3D blocks, delivering consistently stable performance across all nine cases.
Morphology metric sets that primarily focus on density-related or shape-related indicators are inadequate for representing complex urban form, often leading to noticeable discrepancies in retrieval results.
Focusing on density-related indicators, Spacemate can capture block density but cannot capture height variations and intricate building texture.
While BlockShape may still be useful in cases where the urban form is strongly correlated with block shape (e.g., case 1).
OneBMC, as a condensed version of AllBlockMetric, offers a more comprehensive representation than Spacemate and BlockShape. However, it has limitations in ensuring consistent retrieval, particularly in capturing irregular block shapes and more nuanced morphological characteristics.

\section{Discussion}
\label{sec: discuss}
\subsection{Contributions}
Urban morphology indicators have a combinatorial effect in representing urban form. The results demonstrate that our study provides innovative approaches that enable the bi-directional bridging between 3D block-scale urban models and morphology metrics. 
Further, our study provides approaches for evaluating the effectiveness of different morphology metric sets in representing 3D urban form characteristics. The results also show that our proposed AllBlockMetric can well represent 3D city block characteristics in terms of prototype similarity and morphological diversity. 
We outline our contributions:
\begin{enumerate}
    \item Our methodology offers a systematic way for deriving morphology metrics from 3D urban models, facilitating both the characterisation and generation of urban form.
    It enables the flexible retrieval of a diverse range of 3D models with similar urban morphology based on given morphology metrics derived from 3D block-scale models. Hence it links urban form and morphology metrics bi-directionally. 
    The capability allows for the integration of performance evaluation with urban form generation as the morphology metric set includes UMI for performance evaluation.
    It contributes to advance the automation of performance-driven computational urban design.

    \item We extend morphology metric sets to better represent urban morphology, transitioning from 2D to 3D. 
    The vertical dimension indicators represent not only height and its derivatives but also aggregated relationships between vertical and horizontal dimensions, such as GWHR and CAR. 
    We propose an innovative SOMEncoding method to encode morphology metric values based on the trained SOM, allowing for a more context-based representation of 3D model characteristics. 
    As a result, our approach effectively represents 3D city block models.
    
    \item Our systematic approaches offer a comparative method to evaluate block-scale morphology metrics for representing urban form characteristics. 
    Based on the aforementioned metric-to-form model, the effectiveness of metric sets in representing urban form can be evaluated comparatively. 
    By comparing our selected morphology metric sets, we found that AllBlockMetric represents 3D city blocks in NYC effectively, making it a valuable finding for supporting further urban planning and design digitalization.

\end{enumerate}

Our proposed methodology and framework using systematic and comparative approaches can benefit both researchers and urban design practitioners, allowing for flexible application based on their specific needs.
For researchers, this framework enables the identification and selection of appropriate morphology metric sets for further urban form studies and form-performance coupled evaluations at varying levels of detail. 
Our findings advance the integration of urban form and performance optimisation in CUD, promoting the development of environmentally sustainable urban forms within the cityscape.
For urban designers, this approach supports the development of more precise form-based codes \citep{garde2015form-basedcode, chen2016designdimension}, enhancing the understanding of city structure and providing detailed guidelines for regulating urban form development across cities. 
The integration of measurable form-based codes with tangible built forms offers an additional governance and management perspective, complementing governance based on land use and census tract maps \citep{li2024characterizing}.
Our study extends beyond simplified typological categorisation by introducing a systematic approach that preserves urban form complexity. Hence, when applied to specific cities, this methodology can be generalised and help uncover unique spatial patterns, offering a valuable resource for addressing design challenges through built-form and environmental analysis.


\subsection{Limitations and future work}

We discuss the limitations and future directions of this study from the following perspectives. 
First, while AllBlockMetric effectively captures comprehensive urban form characteristics, as demonstrated in the case retrieval experiments and our internal testing, there remains a possibility that the metric set is incomplete or context-dependent. Since the validation was conducted within the New York City (NYC) context, certain nuances may not have been fully addressed. Future studies could explore incorporating additional indicators to better accommodate specific needs.
In this study, 15 morphology indicators representing four characteristic types were selected based on the literature. Our proposed framework is generalisable to any OSM datasets from other cities. However, when extending this to other cities, it is essential to consider local urban characteristics and planning contexts.
To apply this approach to different urban spatial patterns in other cities or countries, it is recommended to reassess the metric set’s performance and, if necessary, integrate additional indicators or refine the framework to ensure broader applicability.

Secondly, while we examined and compared the performance of four morphology metric sets, the influence of individual indicators on characterisation performance, as well as the possibility of developing a more comprehensive metric set for specific urban contexts, has not been fully explored.
Future research could extend this study by incorporating a systematic workflow to further assess the strengths and weaknesses of each indicator, for example, by comparing typical urban configurations with urban design principles in urban contexts. This would enable the identification of key influential indicators, facilitating the development of a refined and optimised morphology metric set for specific urban contexts.

The third limitation concerns data volume and quality. While our methodology effectively characterises and generates urban models based on our morphology metrics, it operates as a retrieval-based generation method. As a result, model performance require a large volume of training data, and a limited dataset may constrain its effectiveness.
In this study, we utilised over 14,000 models, ensuring that certain urban form types had a sufficient number of samples, leading to convincing results. However, when applying this framework to a smaller dataset, it may be necessary to incorporate data augmentation techniques or complement the approach with image-based machine learning methods \citep{fan2021highspatial-resolution, cai2021urbanmorphologyfeatureextraction, dong2019newquantitative} to enhance urban form characterisation.

Fourth, many of the UMIs that constitute our morphology metric sets are derived from urban form indicators primarily used for urban performance analysis, such as microclimate assessment. Although this approach can support performance-driven urban form generation, a case study on performance-driven design is helpful to validate and strengthen our methodology. 
Applying our approaches to retrieved 3D urban form models based on optimised morphology metrics using urban effect evaluation models (e.g., CFD models) \citep{burian2002morphologicalanalysis, da2021blocktypologyonpollutant}, could further demonstrate its practical applicability.
For instance, by simulating and storing the urban performance of these blocks based on the morphology metrics, we can further provide improved urban form models with better performance profiles according to the input urban form types.
In that way, urban performance implications of different urban forms can be further investigated, hence achieving a comprehensive digital link between urban form generation and urban impact analysis. 
Integrating energy simulation into the framework can provide urban designers and planners with tools to assess the environmental impacts of their designs in real-time, fostering sustainable and efficient urban development.

\section{Conclusion}
\label{sec:conclusion}
The accelerating digitalisation in urban planning and design underscores the importance of bi-directional links between intangible urban form complexity and tangible representation in digital models. 
This work, focusing on 3D block-scale urban form and their morphology metrics, presents approaches for evaluating morphology metric sets and associating them with 3D models. 
Our study offers great potential to bridge the generation of improved urban form and optimised morphology metrics in performance-driven CUD.
We proposed a methodology for establishing bi-directional links between morphology metrics and 3D urban form and evaluated the performance of these morphology metric sets. 
Our methodology has innovative contributions in encoding and evaluating morphology metric sets in effectively representing 3D urban forms, and digitally coupling morphology metrics and 3D urban models. 

First, we extracted UMIs from block-scale 3D urban models, including UMIs that describe height, surface ratio, block content, and 2D block shape characteristics. 
The results demonstrate that our morphology metrics represented well the vertical dimensions of the complex 3D blocks.
The 3D models were then clustered based on these morphology metrics using self-organizing maps (SOMs). 
We utilised SOMs not only for clustering but also for deeper encoding of the morphology metrics. 
We leverage the trained SOMs as advanced contexts to further encode the morphology metric sets, hence better representing urban form characteristics in a context-aware manner.

Second, by integrating case retrieval techniques, our approach effectively associates morphology metrics with 3D blocks.
Using the SOMEncoding vectors, case retrieval results demonstrate high consistency in prototype similarity and morphological diversity, effectively addressing the representation of complex urban form. 
By using this approach as a bridge, we can translate our target urban form and morphology metrics into high-level quantifications, and then, conversely, generate more urban models that are similar to the target and maintain the diversity.

Third, our methodology provides a comparative approach to evaluating specific morphology metric sets in representing targeted 3D urban models. 
Using our metric-to-form approach, we can evaluate the performance of morphology metric sets based on the retrieved 3D models. 
In our case of NYC, by comparing four morphology metrics, our AllBlockMetric effectively represents block-scale 3D models, and the OneBMC can be an alternitive for needs of reducing computational costs for large-scale applications.
This approach is flexible and general, making it easily applicable for other evaluation purposes or datasets.

In summary, in this study, our approach provides a general and flexible way to formulate and evaluate morphology metric sets for 3D block-scale models. Among the tested morphology metric sets, AllBlockMetric is a computable metric based on which numerous urban form that are similar in prototype and rich in morphology can be retrieved.
We establish a systematic methodology that simultaneously enables the characterisation, generation, and performance evaluation of urban form. This methodology supports the derivation of morphology metrics from 3D models and the generation of improved urban form with enhanced performance evaluation.
The proposed methodology establishes a bi-directional link between complex urban form with morphology metrics, hence it can drive a seamless integration between urban form generation and optimisation in performance-driven CUD, towards sustainable urban design and planning.

\section*{Acknowledgements}
Part of this research was conducted at the Future Cities Lab Global at Singapore-ETH Centre. Future Cities Lab Global is supported and funded by the National Research Foundation, Prime Minister’s Office, Singapore under its Campus for Research Excellence and Technological Enterprise (CREATE) programme and ETH Zurich (ETHZ). Part of this research was funded by the National Natural Science Foundation of China (NSFC) project (No.52378008).

\section*{CRediT author statement}

\textbf{Chenyi Cai}: Conceptualization, Methodology, Data curation, Formal analysis, Investigation, Visualization, Writing – original draft, Writing – review \& editing. 
\textbf{Biao Li}: Conceptualization, Methodology, Supervision, Writing – review \& editing, Funding acquisition.
\textbf{Qiyan Zhang}: Visualization, Formal analysis, Writing – review \& editing.
\textbf{Xiao Wang}: Writing – review \& editing.
\textbf{Filip Biljecki}: Writing – review \& editing, Funding acquisition.
\textbf{Pieter Herthogs}: Conceptualization, Writing – review \& editing, Funding acquisition.

 \bibliographystyle{elsarticle-harv} 
 \bibliography{cas-refs}

\begin{thebibliography}{95}
\expandafter\ifx\csname natexlab\endcsname\relax\def\natexlab#1{#1}\fi
\providecommand{\url}[1]{\texttt{#1}}
\providecommand{\href}[2]{#2}
\providecommand{\path}[1]{#1}
\providecommand{\DOIprefix}{doi:}
\providecommand{\ArXivprefix}{arXiv:}
\providecommand{\URLprefix}{URL: }
\providecommand{\Pubmedprefix}{pmid:}
\providecommand{\doi}[1]{\href{http://dx.doi.org/#1}{\path{#1}}}
\providecommand{\Pubmed}[1]{\href{pmid:#1}{\path{#1}}}
\providecommand{\bibinfo}[2]{#2}
\ifx\xfnm\relax \def\xfnm[#1]{\unskip,\space#1}\fi
\bibitem[{Anderson et~al.(1996)Anderson, Kanaroglou and Miller}]{anderson1996urbanformenergyandenviron}
\bibinfo{author}{Anderson, W.P.}, \bibinfo{author}{Kanaroglou, P.S.}, \bibinfo{author}{Miller, E.J.}, \bibinfo{year}{1996}.
\newblock \bibinfo{title}{Urban form, energy and the environment: a review of issues, evidence and policy}.
\newblock \bibinfo{journal}{Urban studies} \bibinfo{volume}{33}, \bibinfo{pages}{7--35}.
\bibitem[{Batty(1976)}]{batty1976urbanModeling}
\bibinfo{author}{Batty, M.}, \bibinfo{year}{1976}.
\newblock \bibinfo{title}{Urban modelling}.
\newblock \bibinfo{publisher}{Cambridge University Press Cambridge}.
\bibitem[{Batty(1991)}]{batty1991generatingurbanforms}
\bibinfo{author}{Batty, M.}, \bibinfo{year}{1991}.
\newblock \bibinfo{title}{Generating urban forms from diffusive growth}.
\newblock \bibinfo{journal}{Environment and Planning A} \bibinfo{volume}{23}, \bibinfo{pages}{511--544}.
\bibitem[{Berghauser~Pont and Haupt(2005)}]{berghauserpontSpacemateDensityTypomorphology2005a}
\bibinfo{author}{Berghauser~Pont, M.}, \bibinfo{author}{Haupt, P.}, \bibinfo{year}{2005}.
\newblock \bibinfo{title}{The {{Spacemate}}: {{Density}} and the typomorphology of the urban fabric}.
\newblock \bibinfo{journal}{Nordic Journal of Architectural Research} \bibinfo{volume}{4}, \bibinfo{pages}{55--68}.
\bibitem[{Berghauser~Pont and Haupt(2021)}]{berghauserpontSpacematrixSpaceDensity2021a}
\bibinfo{author}{Berghauser~Pont, M.}, \bibinfo{author}{Haupt, P.}, \bibinfo{year}{2021}.
\newblock \bibinfo{title}{Spacematrix - {{Space}}, {{Density}} and {{Urban Form}}}.
\bibitem[{Biljecki and Chow(2022)}]{biljeckiGlobalBuildingMorphology2022}
\bibinfo{author}{Biljecki, F.}, \bibinfo{author}{Chow, Y.S.}, \bibinfo{year}{2022}.
\newblock \bibinfo{title}{Global {{Building Morphology Indicators}}}.
\newblock \bibinfo{journal}{Computers, Environment and Urban Systems} \bibinfo{volume}{95}, \bibinfo{pages}{101809}.
\newblock \DOIprefix\doi{10.1016/j.compenvurbsys.2022.101809}.
\bibitem[{Biljecki et~al.(2023)Biljecki, Chow and Lee}]{2023_bae_osm_qa}
\bibinfo{author}{Biljecki, F.}, \bibinfo{author}{Chow, Y.S.}, \bibinfo{author}{Lee, K.}, \bibinfo{year}{2023}.
\newblock \bibinfo{title}{{Quality of crowdsourced geospatial building information: A global assessment of OpenStreetMap attributes}}.
\newblock \bibinfo{journal}{Building and Environment} \bibinfo{volume}{237}, \bibinfo{pages}{110295}.
\bibitem[{Bramley and Power(2009)}]{bramley2009urbanformandsustainability}
\bibinfo{author}{Bramley, G.}, \bibinfo{author}{Power, S.}, \bibinfo{year}{2009}.
\newblock \bibinfo{title}{Urban form and social sustainability: the role of density and housing type}.
\newblock \bibinfo{journal}{Environment and Planning B: Planning and Design} \bibinfo{volume}{36}, \bibinfo{pages}{30--48}.
\bibitem[{Bruyns et~al.(2021)Bruyns, Higgins and Nel}]{bruyns2021urbanvolumetrics}
\bibinfo{author}{Bruyns, G.J.}, \bibinfo{author}{Higgins, C.D.}, \bibinfo{author}{Nel, D.H.}, \bibinfo{year}{2021}.
\newblock \bibinfo{title}{Urban volumetrics: From vertical to volumetric urbanisation and its extensions to empirical morphological analysis}.
\newblock \bibinfo{journal}{Urban Studies} \bibinfo{volume}{58}, \bibinfo{pages}{922--940}.
\bibitem[{Burian et~al.(2002)Burian, Maddula, Velugubantla and Brown}]{burian2002morphologicalanalysis}
\bibinfo{author}{Burian, S.J.}, \bibinfo{author}{Maddula, S.R.K.}, \bibinfo{author}{Velugubantla, S.P.}, \bibinfo{author}{Brown, M.J.}, \bibinfo{year}{2002}.
\newblock \bibinfo{title}{Morphological analyses using 3D building databases: Albuquerque, New Mexico}.
\newblock \bibinfo{type}{Technical Report}. LA-UR-02-6198, Los Alamos National Laboratory, Los Alamos, New Mexico.
\bibitem[{Cai et~al.(2021)Cai, Guo, Zhang, Wang, Li and Tang}]{cai2021urbanmorphologyfeatureextraction}
\bibinfo{author}{Cai, C.}, \bibinfo{author}{Guo, Z.}, \bibinfo{author}{Zhang, B.}, \bibinfo{author}{Wang, X.}, \bibinfo{author}{Li, B.}, \bibinfo{author}{Tang, P.}, \bibinfo{year}{2021}.
\newblock \bibinfo{title}{Urban morphological feature extraction and multi-dimensional similarity analysis based on deep learning approaches}.
\newblock \bibinfo{journal}{Sustainability} \bibinfo{volume}{13}, \bibinfo{pages}{6859}.
\bibitem[{Cai et~al.(2022)Cai, Zaghloul and Li}]{cai2022dataclusteringinUCM}
\bibinfo{author}{Cai, C.}, \bibinfo{author}{Zaghloul, M.}, \bibinfo{author}{Li, B.}, \bibinfo{year}{2022}.
\newblock \bibinfo{title}{Data clustering in urban computational modeling by integrated geometry and imagery features for probabilistic navigation}.
\newblock \bibinfo{journal}{Applied Sciences} \bibinfo{volume}{12}, \bibinfo{pages}{12704}.
\bibitem[{Canuto et~al.(2024)Canuto, Koenig, Chronis, Galanos and Celani}]{canuto2024predictive-performativedesign}
\bibinfo{author}{Canuto, R.}, \bibinfo{author}{Koenig, R.}, \bibinfo{author}{Chronis, A.}, \bibinfo{author}{Galanos, T.}, \bibinfo{author}{Celani, G.}, \bibinfo{year}{2024}.
\newblock \bibinfo{title}{From performative to predictive-performative design: A review of current trends in performance-based design and their impact on urbanism}.
\newblock \bibinfo{journal}{International Journal of Architectural Computing} , \bibinfo{pages}{14780771241281883}.
\bibitem[{Chadzynski et~al.(2022)Chadzynski, Li, Grisiute, Farazi, Lindberg, Mosbach, Herthogs and Kraft}]{chadzynski2022semantic3dcityagents}
\bibinfo{author}{Chadzynski, A.}, \bibinfo{author}{Li, S.}, \bibinfo{author}{Grisiute, A.}, \bibinfo{author}{Farazi, F.}, \bibinfo{author}{Lindberg, C.}, \bibinfo{author}{Mosbach, S.}, \bibinfo{author}{Herthogs, P.}, \bibinfo{author}{Kraft, M.}, \bibinfo{year}{2022}.
\newblock \bibinfo{title}{Semantic 3d city agents—an intelligent automation for dynamic geospatial knowledge graphs}.
\newblock \bibinfo{journal}{Energy and AI} \bibinfo{volume}{8}, \bibinfo{pages}{100137}.
\bibitem[{Chatzipoulka et~al.(2018)Chatzipoulka, Compagnon, Kaempf and Nikolopoulou}]{chatzipoulka2018skyviewfactorforsolaravailability}
\bibinfo{author}{Chatzipoulka, C.}, \bibinfo{author}{Compagnon, R.}, \bibinfo{author}{Kaempf, J.}, \bibinfo{author}{Nikolopoulou, M.}, \bibinfo{year}{2018}.
\newblock \bibinfo{title}{Sky view factor as predictor of solar availability on building fa{\c{c}}ades}.
\newblock \bibinfo{journal}{Solar Energy} \bibinfo{volume}{170}, \bibinfo{pages}{1026--1038}.
\bibitem[{Chen(2016)}]{chen2016designdimension}
\bibinfo{author}{Chen, F.}, \bibinfo{year}{2016}.
\newblock \bibinfo{title}{The design dimension of china's planning system: urban design for development control}.
\newblock \bibinfo{journal}{International Planning Studies} \bibinfo{volume}{21}, \bibinfo{pages}{81--100}.
\bibitem[{Cheshmehzangi and Dawodu(2021)}]{cheshmehzangi2021towards}
\bibinfo{author}{Cheshmehzangi, A.}, \bibinfo{author}{Dawodu, A.}, \bibinfo{year}{2021}.
\newblock \bibinfo{title}{Towards a sustainable energy planning strategy: The utilisation of floor area ratio for residential community planning and design in china}.
\newblock \bibinfo{journal}{Frontiers in Sustainable cities} \bibinfo{volume}{3}, \bibinfo{pages}{687895}.
\bibitem[{Chiaradia(2019)}]{chiaradia2019urbanmophoandform}
\bibinfo{author}{Chiaradia, A.J.}, \bibinfo{year}{2019}.
\newblock \bibinfo{title}{Urban morphology/urban form}.
\newblock \bibinfo{journal}{The Wiley Blackwell Encyclopedia of Urban and Regional Studies} , \bibinfo{pages}{1--6}.
\bibitem[{Choi et~al.(2021)Choi, Nguyen and Makki}]{choi2021designofurbantissueevolutionary}
\bibinfo{author}{Choi, J.}, \bibinfo{author}{Nguyen, P.C.T.}, \bibinfo{author}{Makki, M.}, \bibinfo{year}{2021}.
\newblock \bibinfo{title}{The design of social and cultural orientated urban tissues through evolutionary processes}.
\newblock \bibinfo{journal}{International Journal of Architectural Computing} \bibinfo{volume}{19}, \bibinfo{pages}{331--359}.
\bibitem[{Cohen et~al.(2009)Cohen, Huang, Chen, Benesty, Benesty, Chen, Huang and Cohen}]{cohen2009pearson}
\bibinfo{author}{Cohen, I.}, \bibinfo{author}{Huang, Y.}, \bibinfo{author}{Chen, J.}, \bibinfo{author}{Benesty, J.}, \bibinfo{author}{Benesty, J.}, \bibinfo{author}{Chen, J.}, \bibinfo{author}{Huang, Y.}, \bibinfo{author}{Cohen, I.}, \bibinfo{year}{2009}.
\newblock \bibinfo{title}{Pearson correlation coefficient}.
\newblock \bibinfo{journal}{Noise reduction in speech processing} , \bibinfo{pages}{1--4}.
\bibitem[{Dibble et~al.(2019)Dibble, Prelorendjos, Romice, Zanella, Strano, Pagel and Porta}]{dibble2019originofspacesmorphometrics}
\bibinfo{author}{Dibble, J.}, \bibinfo{author}{Prelorendjos, A.}, \bibinfo{author}{Romice, O.}, \bibinfo{author}{Zanella, M.}, \bibinfo{author}{Strano, E.}, \bibinfo{author}{Pagel, M.}, \bibinfo{author}{Porta, S.}, \bibinfo{year}{2019}.
\newblock \bibinfo{title}{On the origin of spaces: Morphometric foundations of urban form evolution}.
\newblock \bibinfo{journal}{Environment and Planning B: Urban Analytics and City Science} \bibinfo{volume}{46}, \bibinfo{pages}{707--730}.
\bibitem[{Dong et~al.(2019)Dong, Li and Han}]{dong2019newquantitative}
\bibinfo{author}{Dong, J.}, \bibinfo{author}{Li, L.}, \bibinfo{author}{Han, D.}, \bibinfo{year}{2019}.
\newblock \bibinfo{title}{New quantitative approach for the morphological similarity analysis of urban fabrics based on a convolutional autoencoder}.
\newblock \bibinfo{journal}{IEEE Access} \bibinfo{volume}{7}, \bibinfo{pages}{138162--138174}.
\bibitem[{Elzeni et~al.(2022)Elzeni, Elmokadem and Badawy}]{elzeni2022classificationofUM}
\bibinfo{author}{Elzeni, M.}, \bibinfo{author}{Elmokadem, A.}, \bibinfo{author}{Badawy, N.M.}, \bibinfo{year}{2022}.
\newblock \bibinfo{title}{Classification of urban morphology indicators towards urban generation}.
\newblock \bibinfo{journal}{Port-Said Engineering Research Journal} \bibinfo{volume}{26}, \bibinfo{pages}{43--56}.
\bibitem[{Fan et~al.(2021)Fan, Ding, Wu, Ge and Li}]{fan2021highspatial-resolution}
\bibinfo{author}{Fan, Y.}, \bibinfo{author}{Ding, X.}, \bibinfo{author}{Wu, J.}, \bibinfo{author}{Ge, J.}, \bibinfo{author}{Li, Y.}, \bibinfo{year}{2021}.
\newblock \bibinfo{title}{High spatial-resolution classification of urban surfaces using a deep learning method}.
\newblock \bibinfo{journal}{Building and Environment} \bibinfo{volume}{200}, \bibinfo{pages}{107949}.
\bibitem[{Fleischmann et~al.(2021)Fleischmann, Romice and Porta}]{fleischmann2021measuring}
\bibinfo{author}{Fleischmann, M.}, \bibinfo{author}{Romice, O.}, \bibinfo{author}{Porta, S.}, \bibinfo{year}{2021}.
\newblock \bibinfo{title}{Measuring urban form: Overcoming terminological inconsistencies for a quantitative and comprehensive morphologic analysis of cities}.
\newblock \bibinfo{journal}{Environment and Planning B: Urban Analytics and City Science} \bibinfo{volume}{48}, \bibinfo{pages}{2133--2150}.
\bibitem[{Gagne and Andersen(2012)}]{gagne2012generativefacadedesign}
\bibinfo{author}{Gagne, J.}, \bibinfo{author}{Andersen, M.}, \bibinfo{year}{2012}.
\newblock \bibinfo{title}{A generative facade design method based on daylighting performance goals}.
\newblock \bibinfo{journal}{Journal of Building Performance Simulation} \bibinfo{volume}{5}, \bibinfo{pages}{141--154}.
\bibitem[{Galster et~al.(2001)Galster, Hanson, Ratcliffe, Wolman, Coleman and Freihage}]{galster2001wrestlingsprawl}
\bibinfo{author}{Galster, G.}, \bibinfo{author}{Hanson, R.}, \bibinfo{author}{Ratcliffe, M.R.}, \bibinfo{author}{Wolman, H.}, \bibinfo{author}{Coleman, S.}, \bibinfo{author}{Freihage, J.}, \bibinfo{year}{2001}.
\newblock \bibinfo{title}{Wrestling sprawl to the ground: defining and measuring an elusive concept}.
\newblock \bibinfo{journal}{Housing policy debate} \bibinfo{volume}{12}, \bibinfo{pages}{681--717}.
\bibitem[{Garde et~al.(2015)Garde, Kim and Tsai}]{garde2015form-basedcode}
\bibinfo{author}{Garde, A.}, \bibinfo{author}{Kim, C.}, \bibinfo{author}{Tsai, O.}, \bibinfo{year}{2015}.
\newblock \bibinfo{title}{Differences between miami's form-based code and traditional zoning code in integrating planning principles}.
\newblock \bibinfo{journal}{Journal of the American Planning Association} \bibinfo{volume}{81}, \bibinfo{pages}{46--66}.
\bibitem[{Grisiute et~al.(2023)Grisiute, Silvennoinen, Li, Chadzynski, Raubal, Kraft, Von~Richthofen and Herthogs}]{grisiute2023semanticspatialpolicy}
\bibinfo{author}{Grisiute, A.}, \bibinfo{author}{Silvennoinen, H.}, \bibinfo{author}{Li, S.}, \bibinfo{author}{Chadzynski, A.}, \bibinfo{author}{Raubal, M.}, \bibinfo{author}{Kraft, M.}, \bibinfo{author}{Von~Richthofen, A.}, \bibinfo{author}{Herthogs, P.}, \bibinfo{year}{2023}.
\newblock \bibinfo{title}{A semantic spatial policy model to automatically calculate allowable gross floor areas in singapore}, in: \bibinfo{booktitle}{International Conference on Computer-Aided Architectural Design Futures}, \bibinfo{organization}{Springer}. pp. \bibinfo{pages}{455--469}.
\bibitem[{Hang et~al.(2012)Hang, Li, Sandberg, Buccolieri and Di~Sabatino}]{hang2012influenceofbuildingheight}
\bibinfo{author}{Hang, J.}, \bibinfo{author}{Li, Y.}, \bibinfo{author}{Sandberg, M.}, \bibinfo{author}{Buccolieri, R.}, \bibinfo{author}{Di~Sabatino, S.}, \bibinfo{year}{2012}.
\newblock \bibinfo{title}{The influence of building height variability on pollutant dispersion and pedestrian ventilation in idealized high-rise urban areas}.
\newblock \bibinfo{journal}{Building and Environment} \bibinfo{volume}{56}, \bibinfo{pages}{346--360}.
\bibitem[{Herfort et~al.(2023)Herfort, Lautenbach, Porto~de Albuquerque, Anderson and Zipf}]{Herfort2023}
\bibinfo{author}{Herfort, B.}, \bibinfo{author}{Lautenbach, S.}, \bibinfo{author}{Porto~de Albuquerque, J.}, \bibinfo{author}{Anderson, J.}, \bibinfo{author}{Zipf, A.}, \bibinfo{year}{2023}.
\newblock \bibinfo{title}{A spatio-temporal analysis investigating completeness and inequalities of global urban building data in openstreetmap}.
\newblock \bibinfo{journal}{Nature Communications} \bibinfo{volume}{14}.
\newblock \URLprefix \url{http://dx.doi.org/10.1038/s41467-023-39698-6}.
\bibitem[{Herold et~al.(2002)Herold, Scepan and Clarke}]{herold2002structureandchangesoflanduse}
\bibinfo{author}{Herold, M.}, \bibinfo{author}{Scepan, J.}, \bibinfo{author}{Clarke, K.C.}, \bibinfo{year}{2002}.
\newblock \bibinfo{title}{The use of remote sensing and landscape metrics to describe structures and changes in urban land uses}.
\newblock \bibinfo{journal}{Environment and planning A} \bibinfo{volume}{34}, \bibinfo{pages}{1443--1458}.
\bibitem[{Jiang et~al.(2024)Jiang, Ma, Webster, Chiaradia, Zhou, Zhao and Zhang}]{jiang2024generativeurbandesignreview}
\bibinfo{author}{Jiang, F.}, \bibinfo{author}{Ma, J.}, \bibinfo{author}{Webster, C.J.}, \bibinfo{author}{Chiaradia, A.J.}, \bibinfo{author}{Zhou, Y.}, \bibinfo{author}{Zhao, Z.}, \bibinfo{author}{Zhang, X.}, \bibinfo{year}{2024}.
\newblock \bibinfo{title}{Generative urban design: A systematic review on problem formulation, design generation, and decision-making}.
\newblock \bibinfo{journal}{Progress in planning} \bibinfo{volume}{180}, \bibinfo{pages}{100795}.
\bibitem[{Jung and Yoon(2021)}]{jung2021analysis}
\bibinfo{author}{Jung, S.}, \bibinfo{author}{Yoon, S.}, \bibinfo{year}{2021}.
\newblock \bibinfo{title}{Analysis of the effects of floor area ratio change in urban street canyons on microclimate and particulate matter}.
\newblock \bibinfo{journal}{Energies} \bibinfo{volume}{14}, \bibinfo{pages}{714}.
\bibitem[{K{\"a}mpf and Robinson(2009)}]{kampf2009optimisationofurbanenergy}
\bibinfo{author}{K{\"a}mpf, J.H.}, \bibinfo{author}{Robinson, D.}, \bibinfo{year}{2009}.
\newblock \bibinfo{title}{Optimisation of urban energy demand using an evolutionary algorithm}, in: \bibinfo{booktitle}{Proceedings of the Eleventh International IBPSA Conference}, pp. \bibinfo{pages}{668--673}.
\bibitem[{K{\"a}mpf and Robinson(2010)}]{kampf2010optimisationofbuildingform}
\bibinfo{author}{K{\"a}mpf, J.H.}, \bibinfo{author}{Robinson, D.}, \bibinfo{year}{2010}.
\newblock \bibinfo{title}{Optimisation of building form for solar energy utilisation using constrained evolutionary algorithms}.
\newblock \bibinfo{journal}{Energy and Buildings} \bibinfo{volume}{42}, \bibinfo{pages}{807--814}.
\bibitem[{Karimimoshaver et~al.(2021)Karimimoshaver, Khalvandi and Khalvandi}]{karimimoshaver2021effectofUM}
\bibinfo{author}{Karimimoshaver, M.}, \bibinfo{author}{Khalvandi, R.}, \bibinfo{author}{Khalvandi, M.}, \bibinfo{year}{2021}.
\newblock \bibinfo{title}{The effect of urban morphology on heat accumulation in urban street canyons and mitigation approach}.
\newblock \bibinfo{journal}{Sustainable Cities and Society} \bibinfo{volume}{73}, \bibinfo{pages}{103127}.
\bibitem[{Kedron et~al.(2019)Kedron, Zhao and Frazier}]{kedron20193DSpatialMetricsForObjects}
\bibinfo{author}{Kedron, P.}, \bibinfo{author}{Zhao, Y.}, \bibinfo{author}{Frazier, A.E.}, \bibinfo{year}{2019}.
\newblock \bibinfo{title}{Three dimensional (3d) spatial metrics for objects}.
\newblock \bibinfo{journal}{Landscape Ecology} \bibinfo{volume}{34}, \bibinfo{pages}{2123--2132}.
\bibitem[{Koenig et~al.(2020)Koenig, Miao, Aichinger, Knecht and Konieva}]{koenig2020integratinganalysisanddesign}
\bibinfo{author}{Koenig, R.}, \bibinfo{author}{Miao, Y.}, \bibinfo{author}{Aichinger, A.}, \bibinfo{author}{Knecht, K.}, \bibinfo{author}{Konieva, K.}, \bibinfo{year}{2020}.
\newblock \bibinfo{title}{Integrating urban analysis, generative design, and evolutionary optimization for solving urban design problems}.
\newblock \bibinfo{journal}{Environment and Planning B: Urban Analytics and City Science} \bibinfo{volume}{47}, \bibinfo{pages}{997--1013}.
\bibitem[{Kohonen(1982)}]{kohonen1982self}
\bibinfo{author}{Kohonen, T.}, \bibinfo{year}{1982}.
\newblock \bibinfo{title}{Self-organized formation of topologically correct feature maps}.
\newblock \bibinfo{journal}{Biological cybernetics} \bibinfo{volume}{43}, \bibinfo{pages}{59--69}.
\bibitem[{Kohonen(2013)}]{kohonen2013essentials}
\bibinfo{author}{Kohonen, T.}, \bibinfo{year}{2013}.
\newblock \bibinfo{title}{Essentials of the self-organizing map}.
\newblock \bibinfo{journal}{Neural networks} \bibinfo{volume}{37}, \bibinfo{pages}{52--65}.
\bibitem[{Kropf(2009)}]{kropfAspectsUrbanForm2009}
\bibinfo{author}{Kropf, K.}, \bibinfo{year}{2009}.
\newblock \bibinfo{title}{Aspects of urban form}.
\newblock \bibinfo{journal}{Urban Morphology} \bibinfo{volume}{13}, \bibinfo{pages}{105--120}.
\newblock \URLprefix \url{https://journal.urbanform.org/index.php/jum/article/view/3949}, \DOIprefix\doi{10.51347/jum.v13i2.3949}. \bibinfo{note}{number: 2}.
\bibitem[{Kropf(2018)}]{kropf2018handbook}
\bibinfo{author}{Kropf, K.}, \bibinfo{year}{2018}.
\newblock \bibinfo{title}{The handbook of urban morphology}.
\newblock \bibinfo{publisher}{John Wiley \& Sons}.
\bibitem[{Labetski et~al.(2023)Labetski, Vitalis, Biljecki, Arroyo~Ohori and Stoter}]{labetski20233dbuildingmetrics}
\bibinfo{author}{Labetski, A.}, \bibinfo{author}{Vitalis, S.}, \bibinfo{author}{Biljecki, F.}, \bibinfo{author}{Arroyo~Ohori, K.}, \bibinfo{author}{Stoter, J.}, \bibinfo{year}{2023}.
\newblock \bibinfo{title}{3d building metrics for urban morphology}.
\newblock \bibinfo{journal}{International Journal of Geographical Information Science} \bibinfo{volume}{37}, \bibinfo{pages}{36--67}.
\bibitem[{Lee et~al.(2016)Lee, Jeong and Kim}]{lee2016impactofindividualtraits}
\bibinfo{author}{Lee, G.}, \bibinfo{author}{Jeong, Y.}, \bibinfo{author}{Kim, S.}, \bibinfo{year}{2016}.
\newblock \bibinfo{title}{Impact of individual traits, urban form, and urban character on selecting cars as transportation mode using the hierarchical generalized linear model}.
\newblock \bibinfo{journal}{Journal of Asian Architecture and Building Engineering} \bibinfo{volume}{15}, \bibinfo{pages}{223--230}.
\bibitem[{Li et~al.(2025)Li, Cui and Yang}]{li2025multi-scale}
\bibinfo{author}{Li, J.}, \bibinfo{author}{Cui, X.}, \bibinfo{author}{Yang, J.}, \bibinfo{year}{2025}.
\newblock \bibinfo{title}{Multi-scale correlation analysis between geometric parameters and solar radiation in high density urban environment——case study in nanjing}.
\newblock \bibinfo{journal}{Frontiers of Architectural Research} \bibinfo{volume}{14}, \bibinfo{pages}{248--266}.
\bibitem[{Li et~al.(2024)Li, He and Zhao}]{li2024high-densityurbanblockformgeneration}
\bibinfo{author}{Li, J.}, \bibinfo{author}{He, Z.}, \bibinfo{author}{Zhao, B.}, \bibinfo{year}{2024}.
\newblock \bibinfo{title}{A novel method of high-density urban block form generation based on multi-objective solar performance optimization: A case study of nanjing}.
\newblock \bibinfo{journal}{Energy and Buildings} , \bibinfo{pages}{114878}.
\bibitem[{Li and Li(2024)}]{li2024characterizing}
\bibinfo{author}{Li, J.}, \bibinfo{author}{Li, C.}, \bibinfo{year}{2024}.
\newblock \bibinfo{title}{Characterizing urban spatial structure through built form typologies: A new framework using clustering ensembles}.
\newblock \bibinfo{journal}{Land Use Policy} \bibinfo{volume}{141}, \bibinfo{pages}{107166}.
\bibitem[{Li et~al.(2022)Li, You and Ding}]{li2022exploringurbanventilation}
\bibinfo{author}{Li, J.}, \bibinfo{author}{You, W.}, \bibinfo{author}{Ding, W.}, \bibinfo{year}{2022}.
\newblock \bibinfo{title}{Exploring urban space quantitative indicators associated with outdoor ventilation potential}.
\newblock \bibinfo{journal}{Sustainable Cities and Society} \bibinfo{volume}{79}, \bibinfo{pages}{103696}.
\bibitem[{Li and Zeng(2024)}]{liMultidisciplinaryParametersCharacterizing2024}
\bibinfo{author}{Li, K.}, \bibinfo{author}{Zeng, H.}, \bibinfo{year}{2024}.
\newblock \bibinfo{title}{Multidisciplinary parameters for characterizing the {{3D}} urban morphology: {{An}} overview based on the relational perspective}.
\newblock \bibinfo{journal}{Sustainable Cities and Society} , \bibinfo{pages}{105364}\DOIprefix\doi{10.1016/j.scs.2024.105364}.
\bibitem[{Li and Quan(2023)}]{li2023identifyingtypologies}
\bibinfo{author}{Li, N.}, \bibinfo{author}{Quan, S.J.}, \bibinfo{year}{2023}.
\newblock \bibinfo{title}{Identifying urban form typologies in seoul using a new gaussian mixture model-based clustering framework}.
\newblock \bibinfo{journal}{Environment and Planning B: Urban Analytics and City Science} \bibinfo{volume}{50}, \bibinfo{pages}{2342--2358}.
\bibitem[{Liu et~al.(2014)Liu, Fan, Wen, Liang and Wu}]{liu2014impactsof3DUM}
\bibinfo{author}{Liu, S.}, \bibinfo{author}{Fan, X.}, \bibinfo{author}{Wen, Q.}, \bibinfo{author}{Liang, W.}, \bibinfo{author}{Wu, Y.}, \bibinfo{year}{2014}.
\newblock \bibinfo{title}{Simulated impacts of 3d urban morphology on urban transportation in megacities: case study in beijing}.
\newblock \bibinfo{journal}{International Journal of Digital Earth} \bibinfo{volume}{7}, \bibinfo{pages}{470--491}.
\bibitem[{Liu et~al.(2020)Liu, Chen, Li and Chen}]{liu2020characterizing3dresidential}
\bibinfo{author}{Liu, Y.}, \bibinfo{author}{Chen, C.}, \bibinfo{author}{Li, J.}, \bibinfo{author}{Chen, W.Q.}, \bibinfo{year}{2020}.
\newblock \bibinfo{title}{Characterizing three dimensional (3-d) morphology of residential buildings by landscape metrics}.
\newblock \bibinfo{journal}{Landscape Ecology} \bibinfo{volume}{35}, \bibinfo{pages}{2587--2599}.
\bibitem[{Liu et~al.(2017)Liu, Wu and Yu}]{liu2017patternsofairpollution}
\bibinfo{author}{Liu, Y.}, \bibinfo{author}{Wu, J.}, \bibinfo{author}{Yu, D.}, \bibinfo{year}{2017}.
\newblock \bibinfo{title}{Characterizing spatiotemporal patterns of air pollution in china: A multiscale landscape approach}.
\newblock \bibinfo{journal}{Ecological Indicators} \bibinfo{volume}{76}, \bibinfo{pages}{344--356}.
\bibitem[{Lowry and Lowry(2014)}]{lowry2014comparingspatialmeatures}
\bibinfo{author}{Lowry, J.H.}, \bibinfo{author}{Lowry, M.B.}, \bibinfo{year}{2014}.
\newblock \bibinfo{title}{Comparing spatial metrics that quantify urban form}.
\newblock \bibinfo{journal}{Computers, Environment and Urban Systems} \bibinfo{volume}{44}, \bibinfo{pages}{59--67}.
\bibitem[{Lynch and Rodwin(1958)}]{lynch1958theoryofurbanform}
\bibinfo{author}{Lynch, K.}, \bibinfo{author}{Rodwin, L.}, \bibinfo{year}{1958}.
\newblock \bibinfo{title}{A theory of urban form}.
\newblock \bibinfo{journal}{Journal of the American institute of planners} \bibinfo{volume}{24}, \bibinfo{pages}{201--214}.
\bibitem[{Mandi{\'c} and Tepav{\v{c}}evi{\'c}(2015)}]{mandic2015shapegrammarapplicationinUD}
\bibinfo{author}{Mandi{\'c}, M.}, \bibinfo{author}{Tepav{\v{c}}evi{\'c}, B.}, \bibinfo{year}{2015}.
\newblock \bibinfo{title}{Analysis of shape grammar application as a tool for urban design}.
\newblock \bibinfo{journal}{Environment and Planning B: Planning and Design} \bibinfo{volume}{42}, \bibinfo{pages}{675--687}.
\bibitem[{Mashhoodi and Unceta(2024)}]{mashhoodi2024urbanformandLST}
\bibinfo{author}{Mashhoodi, B.}, \bibinfo{author}{Unceta, P.M.}, \bibinfo{year}{2024}.
\newblock \bibinfo{title}{Urban form and surface temperature inequality in 683 european cities}.
\newblock \bibinfo{journal}{Sustainable Cities and Society} , \bibinfo{pages}{105690}.
\bibitem[{Miao et~al.(2018)Miao, Koenig, Knecht, Konieva, Bu{\v{s}} and Chang}]{miao2018computationalurbandesignprototyping}
\bibinfo{author}{Miao, Y.}, \bibinfo{author}{Koenig, R.}, \bibinfo{author}{Knecht, K.}, \bibinfo{author}{Konieva, K.}, \bibinfo{author}{Bu{\v{s}}, P.}, \bibinfo{author}{Chang, M.C.}, \bibinfo{year}{2018}.
\newblock \bibinfo{title}{Computational urban design prototyping: Interactive planning synthesis methods—a case study in cape town}.
\newblock \bibinfo{journal}{International Journal of Architectural Computing} \bibinfo{volume}{16}, \bibinfo{pages}{212--226}.
\bibitem[{Moosavi(2015)}]{moosavi2015computational}
\bibinfo{author}{Moosavi, V.}, \bibinfo{year}{2015}.
\newblock \bibinfo{title}{Computational urban modeling: From mainframes to data streams.}, in: \bibinfo{booktitle}{AAAI Workshop: AI for Cities}.
\bibitem[{Moudon(1997)}]{moudon1997urbanmorphologyasemerging}
\bibinfo{author}{Moudon, A.V.}, \bibinfo{year}{1997}.
\newblock \bibinfo{title}{Urban morphology as an emerging interdisciplinary field}.
\newblock \bibinfo{journal}{Urban morphology} \bibinfo{volume}{1}, \bibinfo{pages}{3--10}.
\bibitem[{O'Neill et~al.(1988)O'Neill, Krummel, Gardner, Sugihara, Jackson, DeAngelis, Milne, Turner, Zygmunt, Christensen et~al.}]{O1988indicesoflandscapepattern}
\bibinfo{author}{O'Neill, R.V.}, \bibinfo{author}{Krummel, J.}, \bibinfo{author}{Gardner, R.e.a.}, \bibinfo{author}{Sugihara, G.}, \bibinfo{author}{Jackson, B.}, \bibinfo{author}{DeAngelis, D.}, \bibinfo{author}{Milne, B.}, \bibinfo{author}{Turner, M.G.}, \bibinfo{author}{Zygmunt, B.}, \bibinfo{author}{Christensen, S.}, et~al., \bibinfo{year}{1988}.
\newblock \bibinfo{title}{Indices of landscape pattern}.
\newblock \bibinfo{journal}{Landscape ecology} \bibinfo{volume}{1}, \bibinfo{pages}{153--162}.
\bibitem[{Pan et~al.(2008)Pan, Zhao, Chen, Liang and Sun}]{pan2008analyzingvariationofbuildingdensity}
\bibinfo{author}{Pan, X.Z.}, \bibinfo{author}{Zhao, Q.G.}, \bibinfo{author}{Chen, J.}, \bibinfo{author}{Liang, Y.}, \bibinfo{author}{Sun, B.}, \bibinfo{year}{2008}.
\newblock \bibinfo{title}{Analyzing the variation of building density using high spatial resolution satellite images: the example of shanghai city}.
\newblock \bibinfo{journal}{Sensors} \bibinfo{volume}{8}, \bibinfo{pages}{2541--2550}.
\bibitem[{Pan{\~a}o et~al.(2008)Pan{\~a}o, Gon{\c{c}}alves and Ferr{\~a}o}]{panao2008optimizationurbanbuilding}
\bibinfo{author}{Pan{\~a}o, M.J.O.}, \bibinfo{author}{Gon{\c{c}}alves, H.J.}, \bibinfo{author}{Ferr{\~a}o, P.M.}, \bibinfo{year}{2008}.
\newblock \bibinfo{title}{Optimization of the urban building efficiency potential for mid-latitude climates using a genetic algorithm approach}.
\newblock \bibinfo{journal}{Renewable Energy} \bibinfo{volume}{33}, \bibinfo{pages}{887--896}.
\bibitem[{Qu and Ma(2023)}]{qu2023blocks}
\bibinfo{author}{Qu, B.}, \bibinfo{author}{Ma, J.}, \bibinfo{year}{2023}.
\newblock \bibinfo{title}{From blocks to cities: Morphology structure rooted in 3d patterns and forming clusters at the block level}.
\newblock \bibinfo{journal}{Frontiers of Architectural Research} \bibinfo{volume}{12}, \bibinfo{pages}{1127--1143}.
\bibitem[{Rakha and Reinhart(2012)}]{rakha2012urbanmodelingwalkability}
\bibinfo{author}{Rakha, T.}, \bibinfo{author}{Reinhart, C.}, \bibinfo{year}{2012}.
\newblock \bibinfo{title}{Generative urban modeling: a design work flow for walkability-optimized cities}.
\newblock \bibinfo{journal}{Proceedings of SimBuild} \bibinfo{volume}{5}, \bibinfo{pages}{255--262}.
\bibitem[{Ratti et~al.(2003)Ratti, Raydan and Steemers}]{ratti2003buildingformenvironmentalperformance}
\bibinfo{author}{Ratti, C.}, \bibinfo{author}{Raydan, D.}, \bibinfo{author}{Steemers, K.}, \bibinfo{year}{2003}.
\newblock \bibinfo{title}{Building form and environmental performance: archetypes, analysis and an arid climate}.
\newblock \bibinfo{journal}{Energy and buildings} \bibinfo{volume}{35}, \bibinfo{pages}{49--59}.
\bibitem[{Rode et~al.(2014)Rode, Keim, Robazza, Viejo and Schofield}]{rode2014citiesandenergy}
\bibinfo{author}{Rode, P.}, \bibinfo{author}{Keim, C.}, \bibinfo{author}{Robazza, G.}, \bibinfo{author}{Viejo, P.}, \bibinfo{author}{Schofield, J.}, \bibinfo{year}{2014}.
\newblock \bibinfo{title}{Cities and energy: urban morphology and residential heat-energy demand}.
\newblock \bibinfo{journal}{Environment and Planning B: Planning and Design} \bibinfo{volume}{41}, \bibinfo{pages}{138--162}.
\bibitem[{Rybarczyk and Wu(2014)}]{rybarczyk2014urbanformbicyclemode}
\bibinfo{author}{Rybarczyk, G.}, \bibinfo{author}{Wu, C.}, \bibinfo{year}{2014}.
\newblock \bibinfo{title}{Examining the impact of urban morphology on bicycle mode choice}.
\newblock \bibinfo{journal}{Environment and Planning B: Planning and Design} \bibinfo{volume}{41}, \bibinfo{pages}{272--288}.
\bibitem[{Scheer(2016)}]{scheer2016epistemologyofUM}
\bibinfo{author}{Scheer, B.C.}, \bibinfo{year}{2016}.
\newblock \bibinfo{title}{The epistemology of urban morphology}.
\newblock \bibinfo{journal}{Urban morphology} \bibinfo{volume}{20}, \bibinfo{pages}{5--17}.
\bibitem[{Schirmer and Axhausen(2016)}]{schirmer2016multiscale}
\bibinfo{author}{Schirmer, P.M.}, \bibinfo{author}{Axhausen, K.W.}, \bibinfo{year}{2016}.
\newblock \bibinfo{title}{A multiscale classification of urban morphology}.
\newblock \bibinfo{journal}{Journal of Transport and Land Use} \bibinfo{volume}{9}, \bibinfo{pages}{101--130}.
\bibitem[{Shen and Ye(2024)}]{shen2024environmentalperformancedrivenurbanmodularhousing}
\bibinfo{author}{Shen, X.}, \bibinfo{author}{Ye, X.}, \bibinfo{year}{2024}.
\newblock \bibinfo{title}{Environmental performance driven optimization of urban modular housing layout in singapore}.
\newblock \bibinfo{journal}{Journal of Asian Architecture and Building Engineering} , \bibinfo{pages}{1--14}.
\bibitem[{Shi et~al.(2017)Shi, Fonseca and Schlueter}]{shi2017reviewsimulationbasedurbanformgeneration}
\bibinfo{author}{Shi, Z.}, \bibinfo{author}{Fonseca, J.A.}, \bibinfo{author}{Schlueter, A.}, \bibinfo{year}{2017}.
\newblock \bibinfo{title}{A review of simulation-based urban form generation and optimization for energy-driven urban design}.
\newblock \bibinfo{journal}{Building and Environment} \bibinfo{volume}{121}, \bibinfo{pages}{119--129}.
\bibitem[{da~Silva et~al.(2021)da~Silva, Reis~Jr, Santos, Goulart and de~Alvarez}]{da2021blocktypologyonpollutant}
\bibinfo{author}{da~Silva, F.T.}, \bibinfo{author}{Reis~Jr, N.C.}, \bibinfo{author}{Santos, J.M.}, \bibinfo{author}{Goulart, E.V.}, \bibinfo{author}{de~Alvarez, C.E.}, \bibinfo{year}{2021}.
\newblock \bibinfo{title}{The impact of urban block typology on pollutant dispersion}.
\newblock \bibinfo{journal}{Journal of Wind Engineering and Industrial Aerodynamics} \bibinfo{volume}{210}, \bibinfo{pages}{104524}.
\bibitem[{Soliman et~al.(2018)Soliman, Mackay, Schmidt, Allan and Wang}]{soliman2018quantifyingBCR}
\bibinfo{author}{Soliman, A.}, \bibinfo{author}{Mackay, A.}, \bibinfo{author}{Schmidt, A.}, \bibinfo{author}{Allan, B.}, \bibinfo{author}{Wang, S.}, \bibinfo{year}{2018}.
\newblock \bibinfo{title}{Quantifying the geographic distribution of building coverage across the us for urban sustainability studies}.
\newblock \bibinfo{journal}{Computers, Environment and Urban Systems} \bibinfo{volume}{71}, \bibinfo{pages}{199--208}.
\bibitem[{Steadman(2014)}]{steadman2014density}
\bibinfo{author}{Steadman, P.}, \bibinfo{year}{2014}.
\newblock \bibinfo{title}{Density and built form: integrating ‘spacemate’with the work of martin and march}.
\newblock \bibinfo{journal}{Environment and Planning B: Planning and Design} \bibinfo{volume}{41}, \bibinfo{pages}{341--358}.
\bibitem[{Stouffs and Rafiq(2015)}]{stouffs2015generativeandevolutionary}
\bibinfo{author}{Stouffs, R.}, \bibinfo{author}{Rafiq, Y.}, \bibinfo{year}{2015}.
\newblock \bibinfo{title}{Generative and evolutionary design exploration}.
\newblock \bibinfo{journal}{AI EDAM} \bibinfo{volume}{29}, \bibinfo{pages}{329--331}.
\bibitem[{Tsai(2005)}]{tsai2005quantifyingurbanform}
\bibinfo{author}{Tsai, Y.H.}, \bibinfo{year}{2005}.
\newblock \bibinfo{title}{Quantifying urban form: compactness versus' sprawl'}.
\newblock \bibinfo{journal}{Urban studies} \bibinfo{volume}{42}, \bibinfo{pages}{141--161}.
\bibitem[{Vermeulen et~al.(2013)Vermeulen, K{\"a}mpf and Beckers}]{vermeulen2013urbanformoptimization}
\bibinfo{author}{Vermeulen, T.}, \bibinfo{author}{K{\"a}mpf, J.H.}, \bibinfo{author}{Beckers, B.}, \bibinfo{year}{2013}.
\newblock \bibinfo{title}{Urban form optimization for the energy performance of buildings using citysim}.
\newblock \bibinfo{journal}{Proceedings of CISBAT 2013 cleantech for smart cities and buildings} , \bibinfo{pages}{915--920}.
\bibitem[{Vesanto and Alhoniemi(2000)}]{vesanto2000clustering}
\bibinfo{author}{Vesanto, J.}, \bibinfo{author}{Alhoniemi, E.}, \bibinfo{year}{2000}.
\newblock \bibinfo{title}{Clustering of the self-organizing map}.
\newblock \bibinfo{journal}{IEEE Transactions on neural networks} \bibinfo{volume}{11}, \bibinfo{pages}{586--600}.
\bibitem[{Wan et~al.(2024)Wan, Du, Yuan, Xu, Tang and Zhang}]{wan2024exploringblockenvironmentalcharacteristics}
\bibinfo{author}{Wan, Y.}, \bibinfo{author}{Du, H.}, \bibinfo{author}{Yuan, L.}, \bibinfo{author}{Xu, X.}, \bibinfo{author}{Tang, H.}, \bibinfo{author}{Zhang, J.}, \bibinfo{year}{2024}.
\newblock \bibinfo{title}{Exploring the influence of block environmental characteristics on land surface temperature and its spatial heterogeneity for a high-density city}.
\newblock \bibinfo{journal}{Sustainable Cities and Society} , \bibinfo{pages}{105973}.
\bibitem[{Wang et~al.(2020)Wang, Song and Tang}]{wang2020generativeurbandesignshapegrammar}
\bibinfo{author}{Wang, X.}, \bibinfo{author}{Song, Y.}, \bibinfo{author}{Tang, P.}, \bibinfo{year}{2020}.
\newblock \bibinfo{title}{Generative urban design using shape grammar and block morphological analysis}.
\newblock \bibinfo{journal}{Frontiers of Architectural Research} \bibinfo{volume}{9}, \bibinfo{pages}{914--924}.
\bibitem[{Wilson et~al.(2019)Wilson, Danforth, Davila and Harvey}]{wilson2019generatemasterplans}
\bibinfo{author}{Wilson, L.}, \bibinfo{author}{Danforth, J.}, \bibinfo{author}{Davila, C.C.}, \bibinfo{author}{Harvey, D.}, \bibinfo{year}{2019}.
\newblock \bibinfo{title}{How to generate a thousand master plans: A framework for computational urban design}.
\newblock \bibinfo{journal}{SimAUD 2019} , \bibinfo{pages}{113--119}.
\bibitem[{Xu and Li(2019)}]{xuRenewalDesignofRoma2019}
\bibinfo{author}{Xu, J.}, \bibinfo{author}{Li, B.}, \bibinfo{year}{2019}.
\newblock \bibinfo{title}{Application of {{Case-Based Methods}} and {{Information Technology}} in {{Urban Design}} - {{The Renewal Design}} of the urban region around {{Roma Railway Station}}}, in: \bibinfo{booktitle}{{{CAADRIA}} 2019: {{Intelligent}} \& {{Informed}}}, \bibinfo{address}{Wellington, New Zealand}. pp. \bibinfo{pages}{625--634}.
\newblock \DOIprefix\doi{10.52842/conf.caadria.2019.1.625}.
\bibitem[{Yang et~al.(2022)Yang, Yang, Zhang, Ma, Zhu and Huang}]{yang2022urbanmorphologicalregionalization}
\bibinfo{author}{Yang, L.}, \bibinfo{author}{Yang, X.}, \bibinfo{author}{Zhang, H.}, \bibinfo{author}{Ma, J.}, \bibinfo{author}{Zhu, H.}, \bibinfo{author}{Huang, X.}, \bibinfo{year}{2022}.
\newblock \bibinfo{title}{Urban morphological regionalization based on 3d building blocks—a case in the central area of chengdu, china}.
\newblock \bibinfo{journal}{Computers, Environment and Urban Systems} \bibinfo{volume}{94}, \bibinfo{pages}{101800}.
\bibitem[{Yi and Kim(2015)}]{yi2015agentbasedgeometryoptimization}
\bibinfo{author}{Yi, Y.K.}, \bibinfo{author}{Kim, H.}, \bibinfo{year}{2015}.
\newblock \bibinfo{title}{Agent-based geometry optimization with genetic algorithm (ga) for tall apartment’s solar right}.
\newblock \bibinfo{journal}{Solar Energy} \bibinfo{volume}{113}, \bibinfo{pages}{236--250}.
\bibitem[{Yin et~al.(2018)Yin, Yuan, Lu, Huang and Liu}]{yin2018urbanformandUHI}
\bibinfo{author}{Yin, C.}, \bibinfo{author}{Yuan, M.}, \bibinfo{author}{Lu, Y.}, \bibinfo{author}{Huang, Y.}, \bibinfo{author}{Liu, Y.}, \bibinfo{year}{2018}.
\newblock \bibinfo{title}{Effects of urban form on the urban heat island effect based on spatial regression model}.
\newblock \bibinfo{journal}{Science of the Total Environment} \bibinfo{volume}{634}, \bibinfo{pages}{696--704}.
\bibitem[{Yu et~al.(2015)Yu, Li, Jia, Zhang and Wang}]{yu2015applicationofmulti-objectiveGA}
\bibinfo{author}{Yu, W.}, \bibinfo{author}{Li, B.}, \bibinfo{author}{Jia, H.}, \bibinfo{author}{Zhang, M.}, \bibinfo{author}{Wang, D.}, \bibinfo{year}{2015}.
\newblock \bibinfo{title}{Application of multi-objective genetic algorithm to optimize energy efficiency and thermal comfort in building design}.
\newblock \bibinfo{journal}{Energy and Buildings} \bibinfo{volume}{88}, \bibinfo{pages}{135--143}.
\bibitem[{Zhang et~al.(2018)Zhang, Dana, Shi, Zhang, Wang, Tyagi and Agrawal}]{zhang2018context}
\bibinfo{author}{Zhang, H.}, \bibinfo{author}{Dana, K.}, \bibinfo{author}{Shi, J.}, \bibinfo{author}{Zhang, Z.}, \bibinfo{author}{Wang, X.}, \bibinfo{author}{Tyagi, A.}, \bibinfo{author}{Agrawal, A.}, \bibinfo{year}{2018}.
\newblock \bibinfo{title}{Context encoding for semantic segmentation}, in: \bibinfo{booktitle}{Proceedings of the IEEE conference on Computer Vision and Pattern Recognition}, pp. \bibinfo{pages}{7151--7160}.
\bibitem[{Zhang et~al.(2019)Zhang, Xu, Shabunko, Tay, Sun, Lau and Reindl}]{zhang2019impacttopologysolarpotential}
\bibinfo{author}{Zhang, J.}, \bibinfo{author}{Xu, L.}, \bibinfo{author}{Shabunko, V.}, \bibinfo{author}{Tay, S.E.R.}, \bibinfo{author}{Sun, H.}, \bibinfo{author}{Lau, S.S.Y.}, \bibinfo{author}{Reindl, T.}, \bibinfo{year}{2019}.
\newblock \bibinfo{title}{Impact of urban block typology on building solar potential and energy use efficiency in tropical high-density city}.
\newblock \bibinfo{journal}{Applied Energy} \bibinfo{volume}{240}, \bibinfo{pages}{513--533}.
\bibitem[{Zhang et~al.(2023)Zhang, Ghosh and Park}]{zhang2023spatialmeasures}
\bibinfo{author}{Zhang, P.}, \bibinfo{author}{Ghosh, D.}, \bibinfo{author}{Park, S.}, \bibinfo{year}{2023}.
\newblock \bibinfo{title}{Spatial measures and methods in sustainable urban morphology: A systematic review}.
\newblock \bibinfo{journal}{Landscape and Urban Planning} \bibinfo{volume}{237}, \bibinfo{pages}{104776}.
\bibitem[{Zhang et~al.(2024)Zhang, Wang, Du, Tian, Jia, Ye, Gao, Kuang and Shi}]{zhang2024reviewurbanformgenerationandoptimization}
\bibinfo{author}{Zhang, X.}, \bibinfo{author}{Wang, X.}, \bibinfo{author}{Du, S.}, \bibinfo{author}{Tian, S.}, \bibinfo{author}{Jia, A.}, \bibinfo{author}{Ye, Y.}, \bibinfo{author}{Gao, N.}, \bibinfo{author}{Kuang, X.}, \bibinfo{author}{Shi, X.}, \bibinfo{year}{2024}.
\newblock \bibinfo{title}{A systematic review of urban form generation and optimization for performance-driven urban design}.
\newblock \bibinfo{journal}{Building and Environment} , \bibinfo{pages}{111269}.
\bibitem[{Zhao and Gou(2023)}]{zhao2023influenceofUMonfacadesolarpotential}
\bibinfo{author}{Zhao, K.}, \bibinfo{author}{Gou, Z.}, \bibinfo{year}{2023}.
\newblock \bibinfo{title}{Influence of urban morphology on facade solar potential in mixed-use neighborhoods: Block prototypes and design benchmark}.
\newblock \bibinfo{journal}{Energy and Buildings} \bibinfo{volume}{297}, \bibinfo{pages}{113446}.
\bibitem[{Zhou et~al.(2022)Zhou, Yuan, Hu, Wei, Dang and Sun}]{zhou2022understandingUMeffectonLST}
\bibinfo{author}{Zhou, L.}, \bibinfo{author}{Yuan, B.}, \bibinfo{author}{Hu, F.}, \bibinfo{author}{Wei, C.}, \bibinfo{author}{Dang, X.}, \bibinfo{author}{Sun, D.}, \bibinfo{year}{2022}.
\newblock \bibinfo{title}{Understanding the effects of 2d/3d urban morphology on land surface temperature based on local climate zones}.
\newblock \bibinfo{journal}{Building and Environment} \bibinfo{volume}{208}, \bibinfo{pages}{108578}.
\bibitem[{Zhou et~al.(2024)}]{zhou2024multi-oboptimisation}
\bibinfo{author}{Zhou, S.}, et~al., \bibinfo{year}{2024}.
\newblock \bibinfo{title}{Multi-objective optimisation design of urban morphology driven by climate adaptation}.
\newblock \bibinfo{journal}{Multi-Objective Optimisation Design of Urban Morphology Driven by Climate Adaptation} .

\end{thebibliography}
 
\newpage
\appendix

\section{Trained SOM visualisations }
\label{sec:soms:appendix}
This appendix is supplementary to section \ref{sec:somresults}, visualising trained SOMs based on different morphology metric sets.

\begin{figure}[h!]
\includegraphics[width=\textwidth]{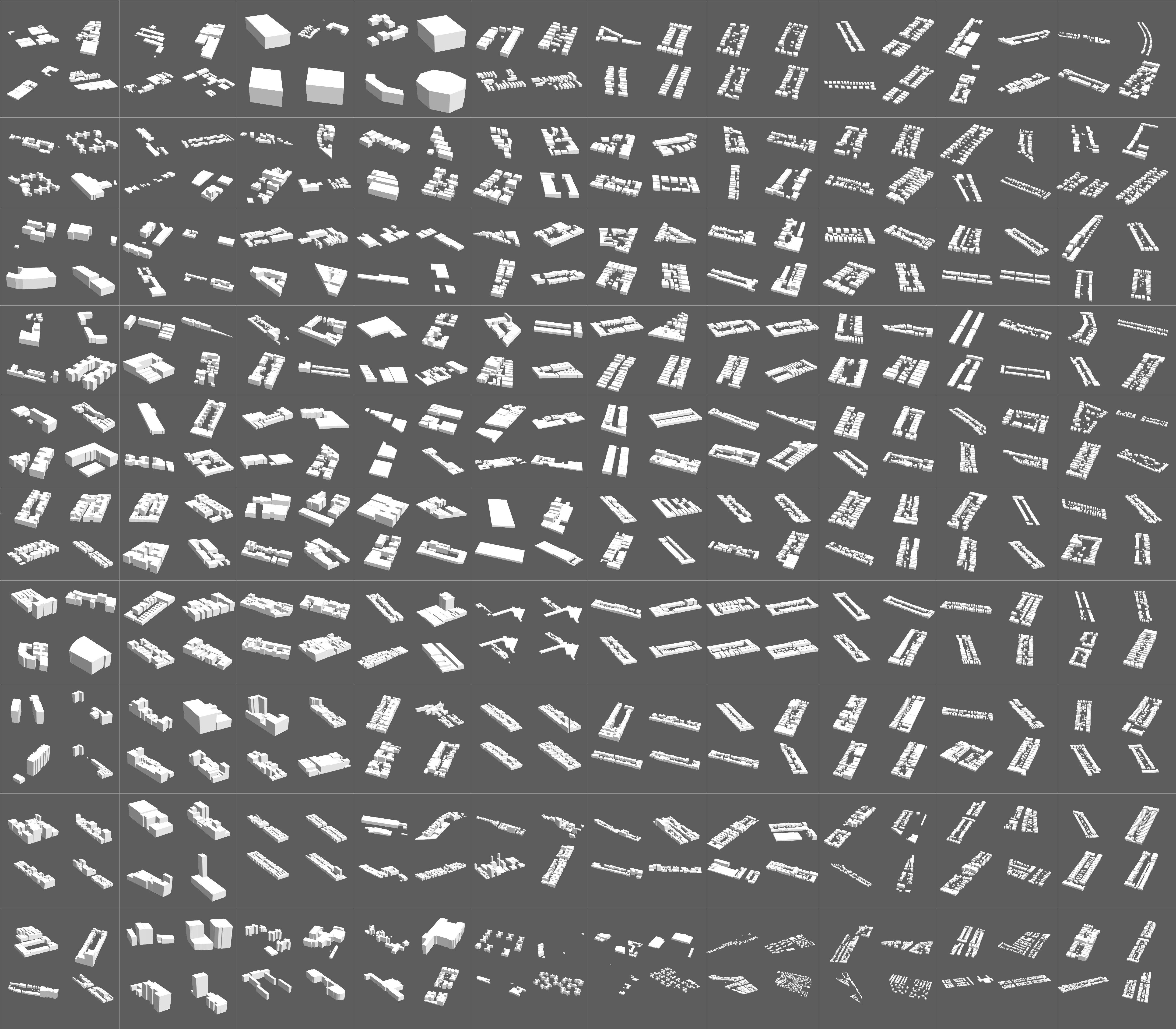}
\centering
\caption{The figure shows the visualisation of the SOM trained by OneBMC set.}
\label{fig:briefblockmetric}
\end{figure}

\begin{figure}[h!]
\includegraphics[width=0.95\textwidth]{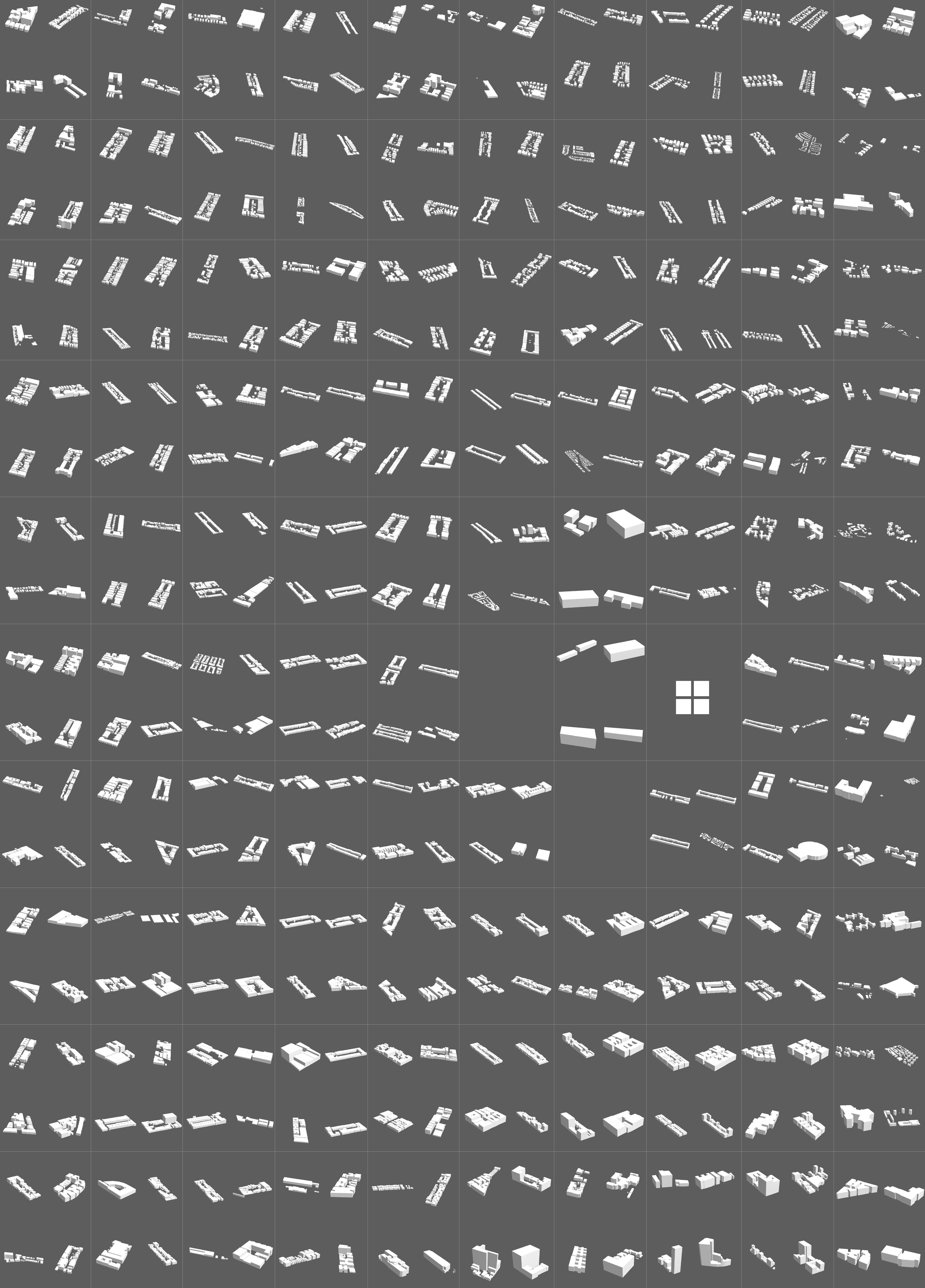}
\centering
\caption{The figure shows the visualisation of the SOM trained by the Spacemate set. The blank grid cells suggest that during the SOM training process, they have no sample assigned as the best matching unit (BMU). This can happen when data points tend to concentrate in specific regions, leaving some neurons unused.
}
\label{fig:spacemate}
\end{figure}

\begin{figure}[h!]
\includegraphics[width=\textwidth]{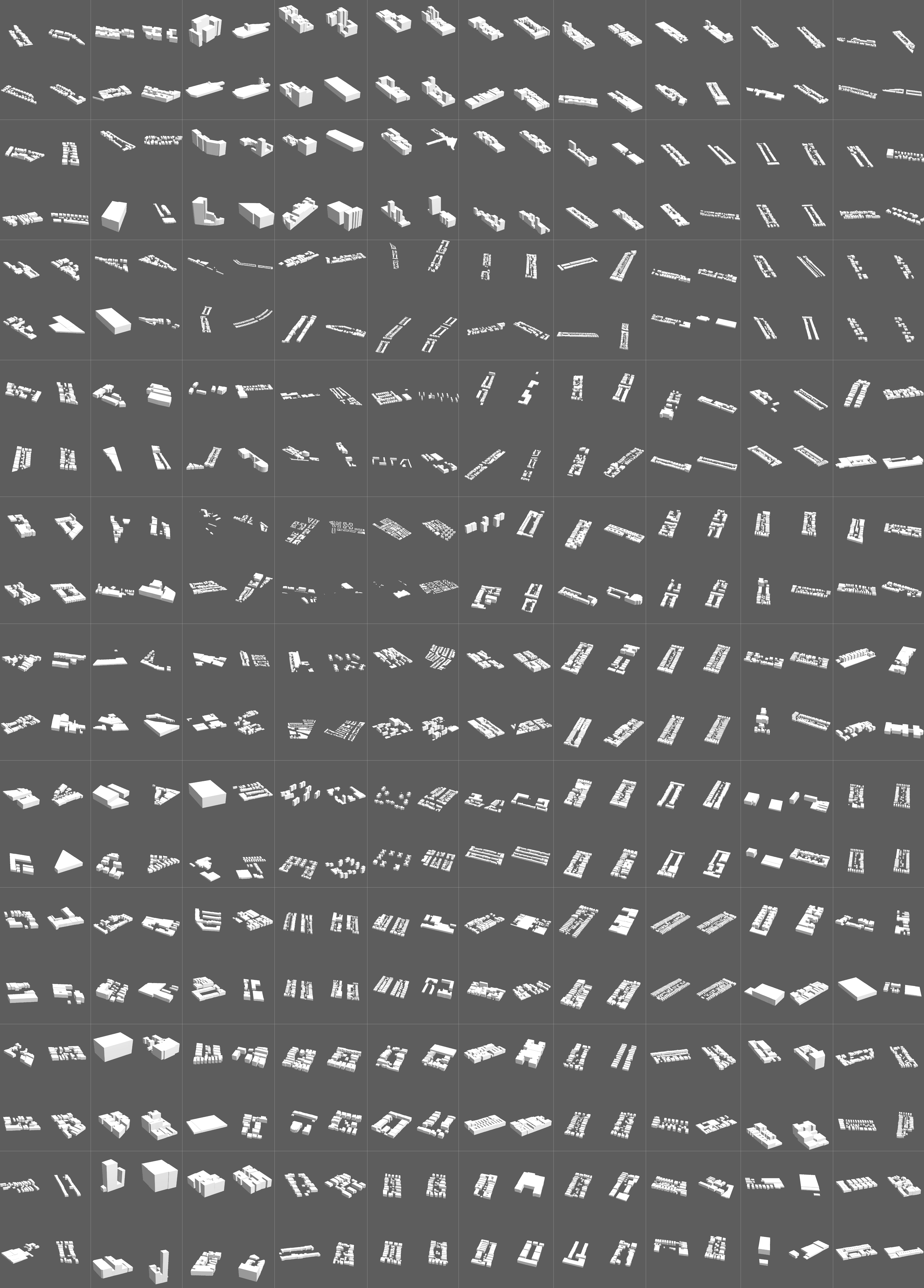}
\centering
\caption{The figure shows the visualisation of the SOM trained by the BlockShape set, visualising 4 models per SOM neuron.}
\label{fig:blockshape}
\end{figure}

\newpage
\section{Evaluating morphology metrics by comparison}
\label{sec:casestudy:appendix}
This appendix is supplementary to section \ref{subsec:evaluate}, showing the urban form retrieval results by comparing 5 morphology metric sets.

\begin{figure}[h!]
\includegraphics[width=\textwidth]{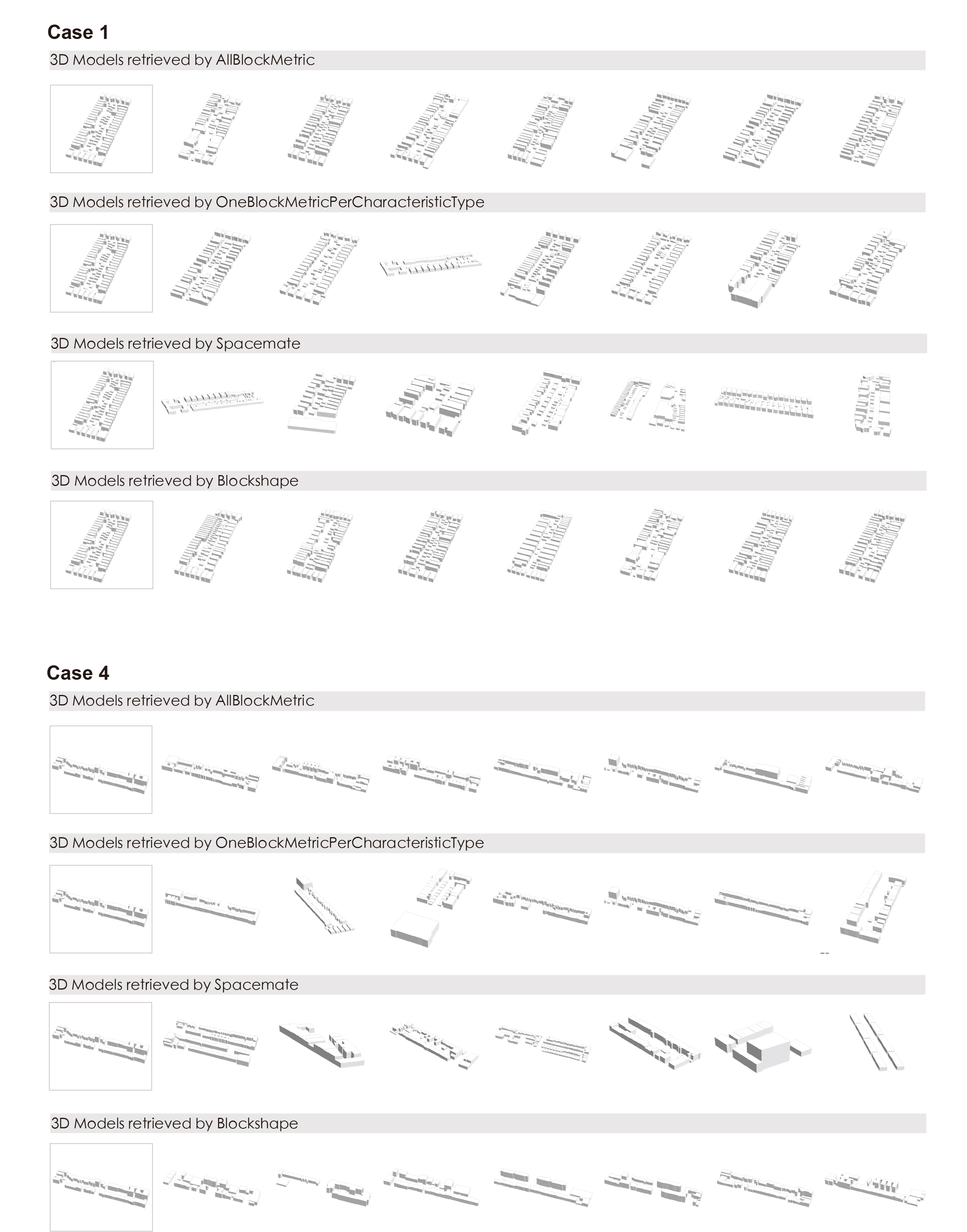}
\centering
\caption{The 3D block-scale models are retrieved based on the distance between SOMEmbedding vectors, comparing the performance of the 4 morphology metric sets for cases 1 and 4.}
\label{fig:c14}
\end{figure}

\begin{figure}[h!]
\includegraphics[width=\textwidth]{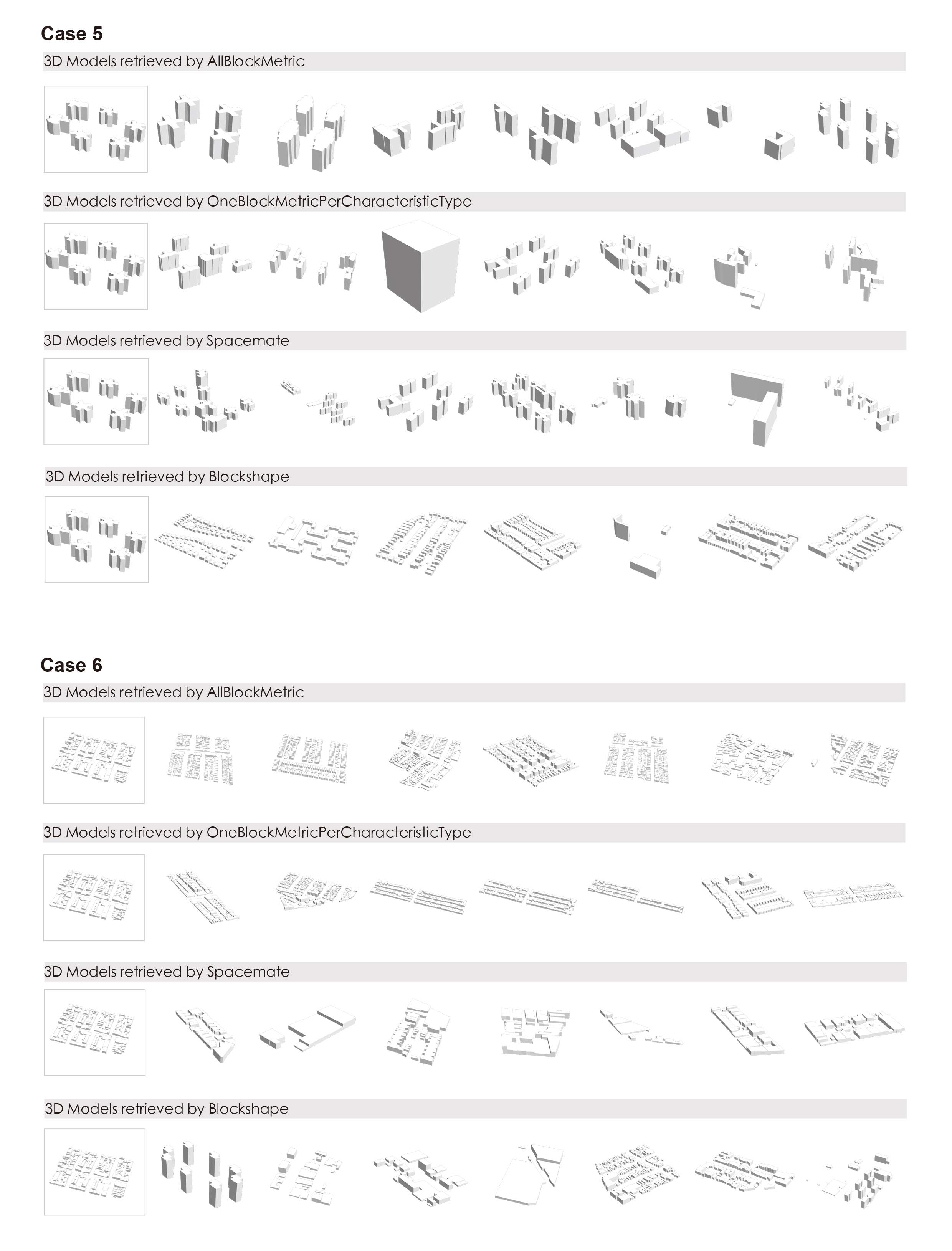}
\centering
\caption{The 3D block-scale models are retrieved based on the distance between SOMEmbedding vectors, comparing the performance of the 4 morphology metric sets for cases 5 and 6.}
\label{fig:c56}
\end{figure}

\begin{figure}[h!]
\includegraphics[width=\textwidth]{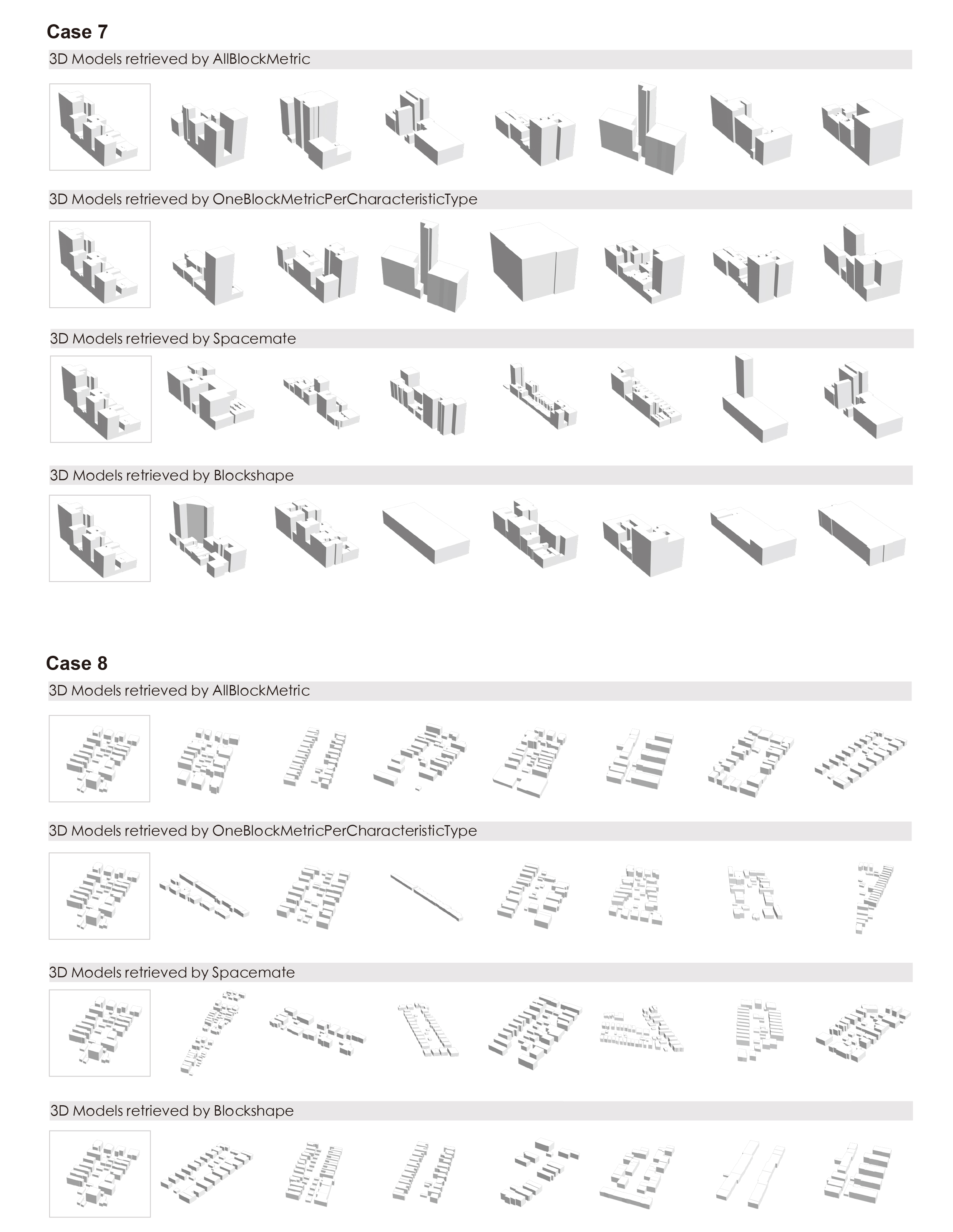}
\centering
\caption{The 3D block-scale models are retrieved based on the distance between SOMEmbedding vectors, comparing the performance of the 4 morphology metric sets for case 7 and 8.}
\label{fig:c78}
\end{figure}

\begin{figure}[h!]
\includegraphics[width=\textwidth]{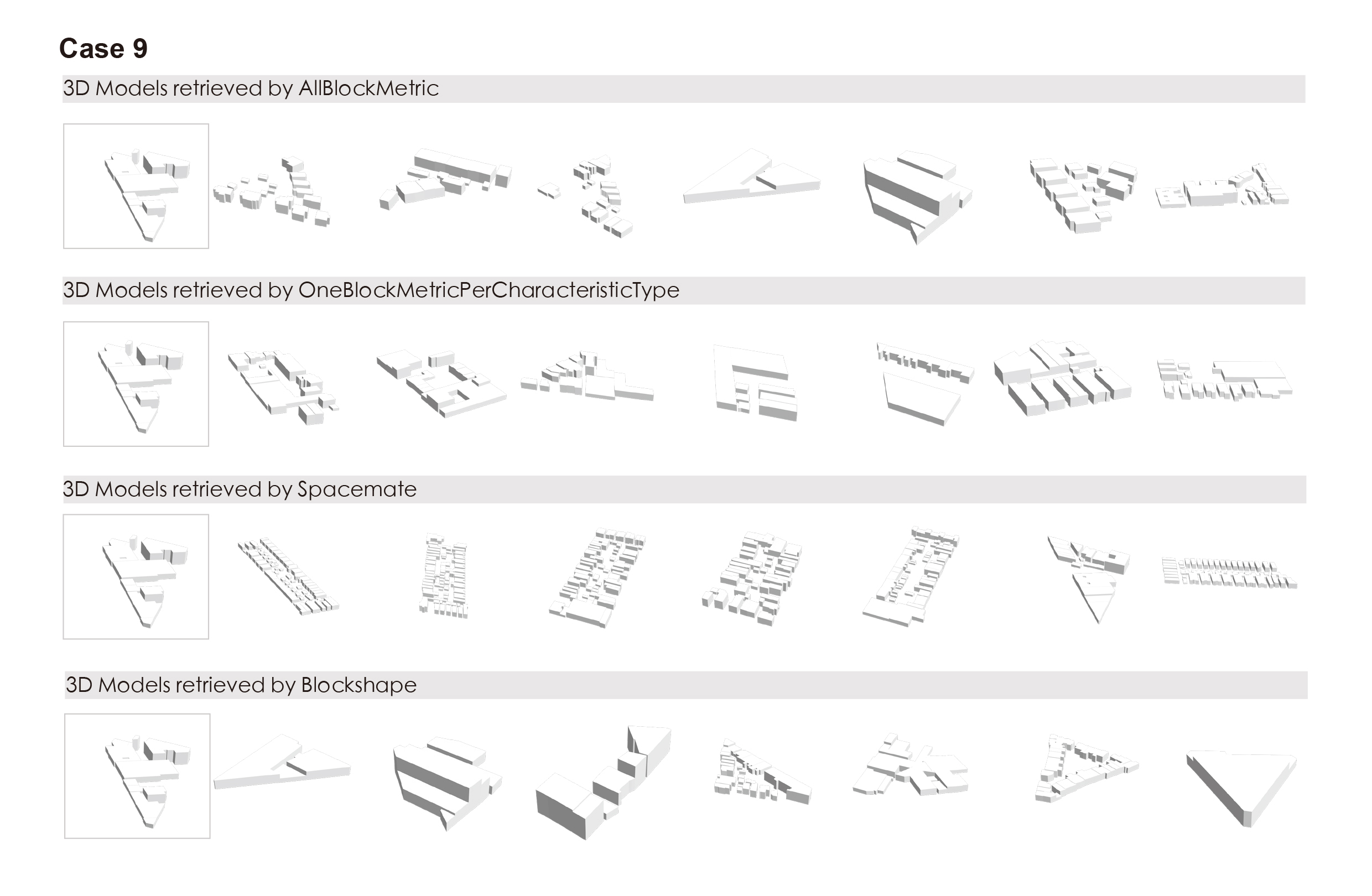}
\centering
\caption{The 3D block-scale models are retrieved based on the distance between SOMEmbedding vectors, comparing the performance of the 4 morphology metric sets for case 9.}
\label{fig:c9}
\end{figure}





\end{document}